\documentclass{ptephy}%%%%where ptephy is the template name

\preprintnumber{XXXX-XXXX} %%% Insert preprint number here

\usepackage{bm}

\begin{document}

\title{Application of a coupled-channel Complex Scaling Method with Feshbach projection to the $K^-pp$ system}

\author{\name{Akinobu Dot\'e}{1,2}, \name{Takashi Inoue}{3}, and \name{Takayuki Myo}{4, 5}
%\thanks{These authors contributed equally to this work}
}
%%%%%%%%%%% The \name command should be used as \name{Insert author name here}{Insert affiliation number here}
%%%%% Please use \thanks for contributed author details

%%%%%%%%%%% The \affil command should be used as \affil{Insert affiliation number here}{Insert author address here}
\address{\affil{1}{KEK Theory Center,
Institute of Particle and Nuclear Studies (IPNS), 
High Energy Accelerator Research Organization (KEK), 1-1 Oho, Tsukuba, Ibaraki, 305-0801, Japan}
\affil{2}{J-PARC Branch, KEK Theory Center, IPNS, KEK,
203-1, Shirakata, Tokai, Ibaraki, 319-1106, Japan}
\affil{3}{Nihon University, College of Bioresource Sciences, Fujisawa 252-0880, Japan}
\affil{4}{General Education, Faculty of Engineering, Osaka Institute of Technology, Osaka 535-8585, Japan}
\affil{5}{Research Center for Nuclear Physics (RCNP), Osaka University, Ibaraki 567-0047, Japan}
\email{dote@post.kek.jp}}

\begin{abstract}%
Kaonic nuclei (nuclear system with anti-kaons) have been an interesting subject in hadron and strange nuclear physics, because the strong attraction between anti-kaon and nucleon might bring exotic properties to that system. In this article, we investigate $K^-pp$ as a prototype of kaonic nuclei. Here, $K^-pp$ is a three-body resonant state in the $\bar{K}NN$-$\pi YN$ coupled channels, where $Y$ represents hyperons $\Lambda$ and $\Sigma$. In order to treat resonant states in a coupled-channel system properly, we propose newly a coupled-channel complex scaling method combined with the Feshbach projection, namely ccCSM+Feshbach method. In this method, the Feshbach projection is realized with help of so-called the extended closure relation held in the complex scaling method, and a complicated coupled-channel problem is reduced to a simple single-channel problem which one can treat easily. First, we confirm that the ccCSM+Feshbach method completely reproduces results of a full coupled-channel calculation in case of two-body $\bar{K}N$-$\pi Y$ system. We then proceed to study of three-body $\bar{K}NN$-$\pi YN$ system, and successfully find solutions of the $K^-pp$ resonance by imposing self-consistency for the complex $\bar{K}N$ energy. Obtained binding energy of $K^-pp$ is well converged around 27 MeV, with an energy-dependent $\bar{K}N$(-$\pi Y$) potential based on the chiral SU(3) theory, independently of ansatz for the self-consistency. This binding energy is small as ones reported in earlier studies based on chiral models. On the other hand, decay width of $K^-pp$ strongly depends on the ansatz. We calculate also the correlation density of $NN$ and $\bar{K}N$ pairs by using the obtained complex-scaled wave function of the $K^-pp$ resonance. Effect of the repulsive core of $NN$ potential is seen in the $NN$ correlation density. In the $\bar{K}N$ correlation density, we can confirm survival of $\Lambda^*$ resonance ($I=0$ $\bar{K}N$ resonance) in the three-body resonance. 
\end{abstract}

\subjectindex{Kaonic nuclei, $K^-pp$ system, coupled-channel problem, resonance, complex scaling method, chiral SU(3) theory}

\maketitle

\section{Introduction}

On hadron and strange nuclear physics, nuclear system with anti-kaons ($\bar{K}$ mesons = ($K^-$, $\bar{K}^0$)) has been a hot issue, since the anti-kaon is expected to cause several interesting phenomena in finite nuclear system due to a strong attraction between anti-kaon and nucleon. In particular, the anti-kaon could be a key to access dense nuclear matter, for which partial restoration of chiral symmetry \cite{ChSymRes:Hatsuda, ChSymRes:Weise} and kaon condensation \cite{Kcond, Kcond-Nstar} have been discussed for a long time.

Both theoretical and experimental studies indicate that the $\bar{K}N$ interaction in the isospin $I=0$ channel is strongly attractive. While, it is known that mass of the excited hyperon $\Lambda(1405)$ cannot be reproduced in a naive quark model with $P$-wave excitation \cite{QM:Isgur}, namely, mass of $\Lambda(1405)$ is predicted about 100 MeV larger than the PDG value exceptionally. Consequently, $\Lambda(1405)$ is considered to be a quasi-bound state of anti-kaon and nucleon since it exists at only $\sim 30$ MeV below $\bar{K}N$ threshold. For example, chiral unitary model \cite{ChU:Review}, based on a meson-baryon dynamics, has successfully explained various properties of the $\Lambda(1405)$. Thus, the $\Lambda(1405)$ is getting recognized as a $\bar{K}N$ quasi-bound state, rather than a genuine three-quark state. Such the attractive nature of the $\bar{K}N$ interaction with a quasi-bound state, is consistent with the repulsive nature of the low-energy $\bar{K}N$ scattering data \cite{Exp:KNscat} and the $1s$-level energy shift of kaonic hydrogen atom \cite{Exp:KpX} which has been updated precisely \cite{Exp:SIDDHARTA}. 

For the $\bar{K}N$ interaction, there are two kinds of potential: one is phenomenological energy-independent potentials ({\it e.g.} Ref. \cite{AY_2002}) and the other is chiral SU(3)-based energy-dependent potentials \cite{ChU:KSW, ChU:OR}. Both are fitted to observables of the $\bar{K}N$ system available at this moment, and are applied and discussed intensively. The former type of potentials is more strongly attractive than the latter type in $\bar{K}N$ sub-threshold region. One study based on a phenomenological $\bar{K}N$ potential \cite{AY_2002} argues  a possibility of so-called deeply bound kaonic nuclei where an anti-kaon is deeply bound in finite nuclei with a binding energy of more than 100 MeV, and such a deeply-bound state could exist as a quasi-stable state since the main decay channel $\pi\Sigma$ is closed. In such a state, nucleons are drawn to the anti-kaon by the strong $\bar{K}N$ attraction, and hence a dense system is generated. In case of light $p$-shell nuclei, studies with antisymmetrized molecular dynamics (AMD) method have shown that the average density amounts to $\sim 4 \rho_0$, when an anti-kaon is added \cite{AMDK}. ($\rho_0$: normal nuclear density, around 0.17 fm$^{-3}$.) In case of medium and heavy nuclei, studies with the relativistic mean field (RMF) approach have been carried out, and it is shown that $K^-$ mesons can create a dense state inside the nucleus \cite{RMF:Mares, RMF:Muto}. Therefore, we can expect kaonic nuclei to be a doorway to dense nuclear matter. 

Thus, kaonic nuclei are considered to be an exotic system involving several interesting aspects from the viewpoint of hadron and nuclear physics. To reveal the nature of kaonic nuclei, a prototype system of kaonic nuclei ``$K^-pp$'' has been studied extensively.\footnote{Actually in the past theoretical studies \cite{Faddeev:Ikeda, Faddeev:Shevchenko, Kpp:AY, Kpp:DHW}, the $\bar{K}NN$-$\pi\Sigma N$-$\pi\Lambda N$ coupled-channel system with the quantum numbers of $J^\pi=0^-$ and $I=1/2$ has been considered. Such a three-body system is denoted symbolically and representatively, as ``$K^-pp$'' in this article.} Since $K^-pp$ is a three-body system composed of a single $K^-$ meson and two protons, various approaches are adopted. As summarized in Ref. \cite{SummaryKpp}, the resulting binding energy and decay width of $K^-pp$ become different depending on the combination of methods (variational approach or Faddeev-AGS) and potentials (phenomenological or chiral-theory based) \cite{Faddeev:Ikeda, Faddeev:Shevchenko, Kpp:AY, Kpp:DHW}. From that time, further studies on $K^-pp$ have been carried out. A variational calculation with the hyperspherical harmonics basis function is reported in Ref.~\cite{Kpp:BGL}, where the result agrees with that of an earlier study of another variational calculation \cite{Kpp:DHW} when they use the same $\bar{K}N$ potential. A Faddeev-AGS calculation with an energy-dependent type of chiral SU(3)-based $\bar{K}N$ potential is reported in Ref.~\cite{Kpp:IKS}, where a $K^-pp$ state with small binding energy and large decay width is obtained similarly to the the variational calculation \cite{Kpp:DHW}. On the other hand, there are several experiments to search for $K^-pp$ states. Actually, the experimental result reported by FINUDA collaboration \cite{Kpp:exp_FINUDA} triggered studies of $K^-pp$, although several questions were casted to their interpretation of the result \cite{Kpp_Criticism:Magas}.  DISTO collaboration \cite{Kpp:exp_DISTO} reported a bump structure found in a $\Lambda p$ invariant-mass distribution from analysis of the past data on a $p+p$ reaction. These two collaborations claim that {\it if the observed state is a $K^-pp$ bound state} the $K^-pp$ is strongly bound with the binding energy of more than 100 MeV, although its decay width is rather different between two. Thus, although much effort have been devoted to the study of $K^-pp$, the definite conclusion has not been achieved yet in both the theoretical and experimental studies. However, in the theoretical side, we have one consensus that the $K^-pp$ exists as a resonance between the $\bar{K}NN$ and $\pi\Sigma N$ thresholds, as commonly reported in all those calculations. 

From those theoretical studies, we believe that the following two ingredients
are important in theoretical studies of the $K^-pp$ system: {\it 1. coupled-channel problem} and {\it 2. resonant state}. We employ a coupled-channel complex scaling method (ccCSM) to study the $K^-pp$, since both the ingredients can be dealt with in this method simultaneously. Here, the complex scaling method (CSM) is an established powerful tool to investigate resonant states, which has already succeeded greatly in the studies of resonant states of stable/unstable nuclei \cite{CSM:Myo, CSM:Myo2}. The CSM has several advantages to investigate finite nuclear systems as follows: First, we can handle resonant states in the same way as bound states, since the resonant wave function in the CSM can be represented by using  only the $L^2$ basis functions such as the Gaussian basis functions which have been often used in bound-state studies. Second, it is straightforward to increase the number of particles in the CSM, which means that we can apply the CSM to various many-body systems. In addition, detailed properties of resonant states can be investigated by analyzing the obtained CSM wave function as usually done for bound states. 

As the first attempt, we have applied the ccCSM to the two-body $\bar{K}N$-$\pi Y$ system in our previous paper~\cite{ccCSM-KN_NPA}. ($Y$ means $\Lambda$ and $\Sigma$ hyperons.) Through the study of scattering states as well as the resonant state $\Lambda(1405)$, we have confirmed that the ccCSM is quite useful to look into such a hadronic system. In that work, we have constructed a $\bar{K}N$-$\pi Y$ potential based on the chiral SU(3) theory, which has a Gaussian form factor in the coordinate space and the energy-dependence. It is shown in studies based on chiral models \cite{ChU:Review}, that $\Lambda(1405)$ should possess so-called a double-pole structure. We reconfirmed such a structure with our Gaussian-type potential. We have successfully identified the lower pole as a broad resonant state, in addition to the higher pole, by using an improved Gaussian basis function in the ccCSM \cite{ccCSM_DP:Dote}. 

Since we have confirmed that the ccCSM is quite effective to the two-body system of $\bar{K}N$-$\pi Y$, we tackle the three-body kaonic nucleus $K^-pp$ in this article. To study the $K^-pp$, normally we have to solve an equation in the coupled-channels $\bar{K}NN$, $\pi\Sigma N$ and $\pi\Lambda N$. But, in this paper we propose a convenient method to reduce such a coupled-channel problem to a single-channel problem, namely, we combine the ccCSM and the Feshbach projection method \cite{Feshbach}. With this method, we can handle the $\bar{K}NN$-$\pi\Sigma N$-$\pi\Lambda N$ complicated system effectively as a simple $\bar{K}NN$ system without loosing effect of the decay to two other open channels. Actually, we study the $K^-pp$ as a Gamow state and obtain the eigenstate as a definite pole on the complex-energy plane. Thus, the $K^-pp$ is treated as a resonant state correctly in this study. In contrast, in the earlier studies of $K^-pp$ with variational approaches \cite{Kpp:DHW, Kpp:BGL}, the $K^-pp$ has been investigated within a bound-state approximation and the decay width is perturbatively estimated with the obtained wave function. 

This article is organized as follows. In the next section, we explain our new method of ccCSM+Feshbach projection in detail and give all tools for the present calculation of the $K^-pp$. In the section 3, we examine our method by solving the two-body $\bar{K}N$-$\pi Y$ system. Main results of this paper, {\it i.e.} application of our method to the three-body $K^-pp$, are shown in the section 4. Section 5 is devoted to summary of the present study and discussion of our future plans.

\section{Methodology \label{Method}}

\subsection{Essence of complex scaling method} 

Here, we give a brief explanation on the usual Complex Scaling Method (CSM) on which the present study is based \cite{CSM:Myo, CSM:Myo2}. In the CSM, Hamiltonian $\hat{H}$ and wave function $|\Phi\rangle$ are transformed with the complex scaling (complex rotation) operator $U(\theta)$ as $\hat{H}_\theta=U(\theta)\hat{H}U^{-1}(\theta)$ and $|\Phi_\theta\rangle = U(\theta)|\Phi\rangle$, respectively. With the complex scaling  the coordinate ${\bm r}$ and the conjugate momentum ${\bm p}$ in the Hamiltonian and wave function are transformed as
\begin{equation}\label{CSM}
\begin{split}
{\bm r} \; \rightarrow \; {\bm r}e^{i\theta}, \quad {\bm p} \; \rightarrow \; {\bm p}e^{-i\theta},
\end{split}
\end{equation}
where the variable $\theta$ is called as the scaling angle. 

In eigenvalues of the complex-scaled Hamiltonian, those of scattering continuum states appear along so-called $2\theta$ line on the complex-energy plane, which satisfies a relation $\tan^{-1}({\rm Im} \, E / {\rm Re} \, E) = -2\theta$. (The variable $E$ means a complex eigen energy with $H_\theta|\Phi_\theta\rangle=E|\Phi_\theta\rangle$.) Namely, they are dependent on the scaling angle. On the other hand, eigenvalues of bound and resonant states are proven to be independent of the scaling angle. In addition, as is easily checked, if we choose appropriate values of the scaling angle $\theta$, wave functions of resonant states are transformed to become square-integrable, which are originally not so. Therefore, the resonant-state wave function, which is complex-scaled, can be expanded with a square-integrable $L^2$ basis function such as Gaussian basis functions, similarly to the bound-state wave functions. 

Due to those nature of the CSM, we can obtain the eigen energies and eigen wave functions of resonant states, by diagonalizing the complex-scaled Hamiltonian with Gaussian basis functions. Detailed explanation on the complex scaling method is summarized in Ref. \cite{CSM:Myo}.

\subsection{Feshbach projection on the coupled-channel Complex Scaling Method \label{Method_Feshbach}}

In the present study, we reduce a coupled-channel problem to a single-channel problem for an economical calculation, based on Feshbach projection method \cite{Feshbach}. In the Feshbach method, a model space ($P$ space) and outer space of the model space ($Q$ space) are assigned with $P+Q=1$ and $PQ=0$. Schr\"odinger equation is given as a coupled equation of wave functions for $P$ and $Q$ spaces as 
\begin{equation}\label{Fesh_cc}
\begin{split}
\left(
\begin{array}{cc}
H_{PP} & V_{PQ} \\
V_{QP} & H_{QQ} 
\end{array}
\right)
\left(
\begin{array}{c}
\Phi_{P}  \\
\Phi_{Q}  
\end{array}
\right)
&= 
E 
\left(
\begin{array}{c}
\Phi_{P}  \\
\Phi_{Q}  
\end{array}
\right), 
\end{split}
\end{equation}
which $\Phi_P$ and $\Phi_Q$ denote $P$- and $Q$-space wave functions, respectively. By the elimination of the $Q$-space wave function, an equation for the $P$-space wave function is derived from Eq. (\ref{Fesh_cc}):   
\begin{equation}\label{Fesh_s1}
\begin{split}
\left\{ H_{PP} \, + \, V_{PQ} G_Q(E) V_{QP} \right\} \Phi_{P} &= E \Phi_{P} \quad {\rm with} \quad G_Q(E) = \frac{1}{E-H_{QQ}}. 
\end{split}
\end{equation}
Since Hamiltonian for the $P$ space $H_{PP}$ is composed of the kinetic energy term $T_P$ and the potential term $V_P$, the above equation can be written as  
\begin{equation}\label{Fesh_s2}
\begin{split}
\left\{ T_P \, + \, U^{eff}_P (E) \right\} \Phi_{P} &= E \Phi_{P} \quad {\rm with} \quad U^{eff}_P (E) = V_P \, + \, V_{PQ} G_Q(E) V_{QP}.  
\end{split}
\end{equation}
Here, the term $U^{eff}_P (E)$ is regarded as an effective potential for the $P$ space. Thus, we obtain a single-channel Schr\"odinger equation for the $P$-space wave function in a formal way. 

In application of Feshbach method to actual studies, the problem is how to represent the $Q$-space Green function, $G_Q(E)$ in Eq. (\ref{Fesh_s1}). We realize the Feshbach method with help of a nature of the complex scaling method (CSM) as follows. It is proven that the closure relation holds in the CSM which includes explicitly resonant states as well as continuum scattering states and bound states \cite{CSM-ECR-proof:Giraud}. ({\it Extended Closure Relation}, ECR, proposed by Berggren \cite{ECR:Berggren}) The ECR is shown to be useful to represent the Green function of a system \cite{CSM-ECR:Myo}. In addition, the ECR is well described approximately with a set of finite number of the complex-scaled eigenstates $\{\phi^\theta_n\}$ which are obtained by the diagonalization of a complex-scaled Hamiltonian $\hat{H}_\theta$ with Gaussian basis functions $\{G_a\}$ \cite{CSM-ECR-Gauss:Suzuki}:  
\begin{equation}\label{ECR}
\begin{split}
\hat{H}_\theta | \phi^\theta_n \rangle \; = \; \epsilon^\theta_n | \phi^\theta_n \rangle 
\quad {\rm with} \quad 
| \phi^\theta_n \rangle \; = \; \sum_{a=1}^M C^{n, \theta}_a | G_a \rangle 
\quad \quad \Longrightarrow \quad
\sum_{n=1}^N | \phi^\theta_n \rangle \langle \tilde{\phi}^\theta_n | \; \simeq \; 1, 
\end{split}
\end{equation}
where $n$ is the state index and complex parameters $\{C^{n, \theta}_a\}$ are determined by a diagonalization of $\hat{H}_\theta$. 

We incorporate the ECR on the $Q$ space into the Feshbach method. 
First, we consider the complex-scaled Green function for the $Q$-space, $G_Q^\theta (E) = U(\theta) G_Q(E) U^{-1}(\theta)$. With the application of the ECR shown in Eq. (\ref{ECR}), it is given approximately as 
\begin{equation}\label{Fesh_GQ}
\begin{split}
G_Q^\theta (E) \; = \;  \frac{1}{E-H_{QQ}^{\theta}}
\quad \simeq \quad \sum_{n=1}^N |\phi^\theta_{Q,n}\rangle 
\frac{1}{E-e^\theta_{Q,n}}
\langle \tilde{\phi}^\theta_{Q,n} |,
\end{split}
\end{equation}
where eigenenergies $\{e^\theta_{Q,n}\}$ and eigenstates $\{|\phi^\theta_{Q,n}\rangle\}$ of the complex-scaled Hamiltonian $H_{QQ}^\theta$ are calculated with Gaussian basis functions $\{G_a\}$. 
By the inverse transformation $U^{-1}(\theta)$, we obtain the non-scaled Green function $G_Q(E)$ from the complex-scaled one $G_Q^\theta (E)$; $G_Q (E) = U^{-1}(\theta) G_Q^\theta(E) U(\theta)$. 
Substituting the obtained $G_Q (E)$ to Eq. (\ref{Fesh_s2}), we can represent the effective $P$-space potential as 
\begin{equation}\label{Fesh_Ueff}
\begin{split}
U^{eff}_P (E) \; = \; V_P \, + \, \sum_{n=1}^N \, U^{-1}(\theta) \, V_{PQ}^{\theta} \; |\phi^\theta_{Q,n}\rangle 
\frac{1}{E-e^\theta_{Q,n}}
\langle \tilde{\phi}^\theta_{Q,n} | \; V_{QP}^\theta \, U(\theta),  
\end{split}
\end{equation}
where $V_{PQ \,(QP)}^\theta = U(\theta) V_{PQ \,(QP)} U^{-1}(\theta)$. 
Since the eigenstates $\{|\phi^\theta_{Q,n}\rangle\}$ are expanded with Gaussian basis function, the effective potential is expressed with Gaussian functions when the original coupled-channel potential is given in a Gaussian form. Therefore, the effective potential $U^{eff}_P (E)$ derived in this way is easily handled in conventional many-body calculations with Gaussian basis functions. 

Thus, we reduce a coupled-channel problem to a single-channel problem with Feshbach projection method which is assisted with a unique nature of the complex scaling method. We call this method as a coupled-channel complex scaling method with Feshbach projection, which is hereafter denoted shortly as {\it ``ccCSM+Feshbach method''}.

\subsection{Hamiltonian and trial wave function for the single $\bar{K}NN$ channel}

In theoretical studies, the $K^-pp$ system is treated as a coupled-channel system of $\bar{K}NN$, $\pi\Sigma N$ and $\pi\Lambda N$, involving quantum numbers 
$J^\pi=0^-$ and $I=1/2$. We apply the ccCSM+Feshbach projection to the $\bar{K}NN$-$\pi YN$ coupled-channel problem to reduce a $\bar{K}NN$ single-channel problem. ($Y=\Lambda$, $\Sigma$)

At first, we consider the two-body system of $\bar{K}N$-$\pi Y$. When we set the $\bar{K}N$ channel to $P$ space and the $\pi Y$ channels to $Q$ space, we can derive an effective $\bar{K}N$ potential for each isospin state ($I=0, 1$) with the ccCSM+Feshbach method: 
\begin{equation}\label{Ueff_KN}
\begin{split}
&U^{eff}_{\bar{K}N(I)} (E_{\bar{K}N}) \; = \; V_{\bar{K}N, \bar{K}N (I)} \, + \, \\
&\quad \quad \sum_{Y(I)} \sum_{n=1}^N \, U^{-1}(\theta_Q) \, V_{\bar{K}N, \pi Y (I)}^{\theta_Q} \; |\phi^{\theta_Q}_{\pi Y (I),n}\rangle 
\frac{1}{E_{\bar{K}N}-e^{\theta_Q}_{\pi Y (I),n}}
\langle \tilde{\phi}^{\theta_Q}_{\pi Y (I),n} | \; V_{\pi Y, \bar{K}N (I)}^{\theta_Q} \, U(\theta_Q),  
\end{split}
\end{equation}
where the index $Y(I)$ indicates $\Sigma$ for $I=0$ and ($\Lambda$, $\Sigma$) for $I=1$. As explained in the previous section, eigenstates and eigenenergies $\{|\phi^{\theta_Q}_{\pi Y(I),n}\rangle, e^{\theta_Q}_{\pi Y (I),n}\}$ are calculated with diagonalization of the complex-scaled Hamiltonian $H_{\pi Y, \, I}^{\theta_Q}$ for each isospin $I$ channel, which is  
\begin{equation}\label{H_piY_rot}
\begin{split}
H_{\pi Y, \, I}^{\theta_Q} \; = \; \sum_{\alpha\,=\,\pi Y(I)} \left( \Delta M_\alpha + \hat{T}_\alpha^{\theta_Q} \right) |\alpha \rangle \langle \alpha| \; + \; \sum_{\alpha, \beta\,=\,\pi Y(I)} \hat{V}_{\alpha \beta}^{(I), {\theta_Q}} \, |\alpha \rangle \langle \beta|,\end{split}
\end{equation}
where the channel indices $\alpha$ and $\beta$ are $\pi\Sigma$ for $I=0$, and 
($\pi\Lambda$, $\pi\Sigma$) for $I=1$. The terms of $\Delta M_\alpha$ and $\hat{T}_\alpha^{\theta_Q}$ are a mass of the $\alpha$ channel measured from the $\bar{K}N$ threshold and the relative kinetic energy term of the $\alpha$ channel which is complex-scaled, respectively. The last term $\hat{V}_{\alpha \beta}^{(I), \theta_Q}$ is a complex-scaled potential coupling between channels $\alpha$ and $\beta$ with isospin $I$. Note that hereafter in this article the variable ``$\theta_Q$'' means the scaling angle which is used to construct an effective potential by the elimination of $Q$-space components as explained in the previous section. 

With the effective $\bar{K}N$ potential $U^{eff}_{\bar{K}N(I)} (E)$ plugged in, a three-body Hamiltonian for the single $\bar{K}NN$ channel is constructed to be \begin{equation}\label{H_KNN}
\begin{split}
\hat{H}_{\bar{K}NN} \; &=  \; \frac{\hat{\bm{p}}_1^2}{2\mu_{NN}}
\; + \; \frac{\hat{\bm{p}}_2^2}{2\mu_{\bar{K}(NN)}} 
\; + \; \hat{V}_{NN}
\; + \; \sum_{i=1,2} \sum_{I=0,1} \hat{U}^{eff}_{(\bar{K}N_i)I} (E_{\bar{K}N}). \end{split}
\end{equation}
The first two terms are kinetic-energy operators with respect to a Jacobi coordinate, ${\bm x}_1 = {\bm r}_{N2}-{\bm r}_{N1}$ and ${\bm x}_2 = {\bm r}_{K}-({\bm r}_{N1}+{\bm r}_{N2})/2$. The term $\hat{V}_{NN}$ is a nucleon-nucleon potential. The last term is the effective $\bar{K}N$ potential for a $\bar{K}N_i$ pair with isospin $I$. Detailed explanation on the $NN$ and $\bar{K}N$ potentials will be given at the beginning of the sections \ref{Sec:Result-2body} and \ref{Sec:Result-3body}. 

A trial wave function of the $\bar{K}NN$ system with quantum numbers 
$(J^\pi, T)=(0^-, 1/2)$ is constructed in the similar way to an earlier study with a variational approach \cite{Kpp:DHW}. Since the spin of the two nucleons are assumed to be zero, the trial wave function consists of two components that satisfy the antisymmetrization for two nucleons. In one component, $NN$ state has even-parity and isospin 1, and in the other component it has odd-parity and isospin 0;  
\begin{equation}\label{Wfunc-KNN}
\begin{split}
|\Phi_{\bar{K}NN} \rangle \; = \; \; \Phi_{\bar{K}NN}^{(+)}({\bm x}_1, {\bm x}_2) \; | S_{NN}=0 \rangle \; |[\bar{K}[NN]_1]_{(T,T_z)=(1/2, 1/2)}\rangle\\
\; + \; \; 
\Phi_{\bar{K}NN}^{(-)}({\bm x}_1, {\bm x}_2) \; | S_{NN}=0 \rangle \, |[\bar{K}[NN]_0]_{(T,T_z)=(1/2, 1/2)}\rangle. 
\end{split}
\end{equation}
In the present study, the spatial part of the wave function $\Phi_{\bar{K}NN}^{(\pm)}({\bm x}_1, {\bm x}_2)$ is expanded with correlated Gaussian basis functions \cite{CG:Suzuki}, so that the $NN$ parity is realized correctly in each part; \begin{equation}\label{Wfunc-KNN-space}
\begin{split}
\Phi_{\bar{K}NN}^{(\pm)}({\bm x}_1, {\bm x}_2) 
\; = \; &\sum_i \, C^{(\pm)}_i \, G^{(\pm)}_i ({\bm x}_1, {\bm x}_2) \\
& {\rm with} \quad G^{(\pm)}_i ({\bm x}_1, {\bm x}_2) \; \equiv \; 
G_i ({\bm x}_1, {\bm x}_2) \, \pm \, G_i (-{\bm x}_1, {\bm x}_2). 
\end{split}
\end{equation}
The variables $\{C^{(\pm)}_i\}$ is complex-valued parameters which are determined by the diagonalization of the complex-scaled Hamiltonian. 
Here, the correlated Gaussian function is 
\begin{equation}\label{Wfunc-CorrG}
\begin{split}
G_i ({\bm x}_1, {\bm x}_2) \; = \; {\cal N}_i \exp [-\tilde{\bm x}^T A_i \, \tilde{\bm x}], 
\end{split}
\end{equation}
where $\tilde{\bm x}^T$ indicates a Jacobi coordinate $({\bm x}_1, {\bm x}_2)$, $A_i$ is a real-symmetric $2\times2$ matrix and ${\cal N}_i$ is a normalization factor. We comment that the basis functions, $G^{(+)}_i ({\bm x}_1, {\bm x}_2)$ and $G^{(-)}_i ({\bm x}_1, {\bm x}_2)$, are even- and odd-parity functions for the exchange of two nucleons, respectively.  

Resonant states of the $\bar{K}NN$ system are obtained in usual way with the complex scaling method. The Hamiltonian for the $\bar{K}NN$ system, $\hat{H}_{\bar{K}NN}$ given in Eq. (\ref{H_KNN}), is complex-scaled with a scaling angle $\theta_P$. The complex-scaled Hamiltonian, $\hat{H}_{\bar{K}NN}^{\theta_P}=U(\theta_P)\,\hat{H}_{\bar{K}NN}\,U^{-1}(\theta_P)$, is diagonalized with the basis functions $\{ G^{(\pm)}_i ({\bm x}_1, {\bm x}_2) \}$ involving spin-isospin wave functions. It is remarked that the scaling angle used to find resonant states of $\bar{K}NN$ system is denoted as ``$\theta_P$'' hereafter, to distinguish from the scaling angle $\theta_Q$ which is used for the construction of the effective potential.

\subsection{Treatment of an energy dependence of the effective potential \label{Sec:SC}}

The effective $\bar{K}N$ potential $U^{eff}_{\bar{K}N(I)} (E_{\bar{K}N})$ which is constructed with the ccCSM+Feshbach has an energy dependence. As shown in Eq. (\ref{Ueff_KN}), the potential depends on a $\bar{K}N$ energy ($E_{\bar{K}N}$) which means the energy of a $\bar{K}N$ system included in a total system that we are considering. In other words, to determine the potential strength we need to know the energy of a $\bar{K}N$ two-body system in the $\bar{K}NN$ three-body system. However, such an energy of a subsystem in a total system cannot be determined uniquely in principle. We deal with the energy dependence of the effective potential, following a procedure proposed in a former study \cite{Kpp:DHW} in which the same issue was considered. 

\begin{figure}[t]
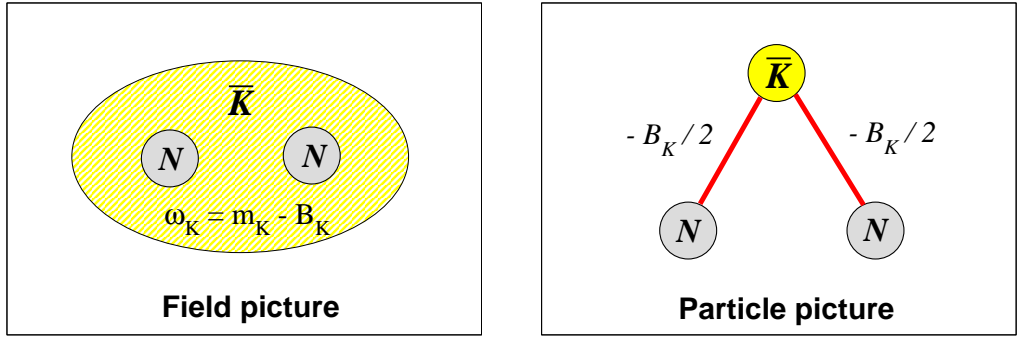

\centering
\includegraphics[width=2.5in]{Fig7-1.eps}
\hspace{0.5cm}
\includegraphics[width=2.5in]{Fig7-2.eps}
\caption{Image of two pictures. (Left) Field picture. (Right) Particle picture. The quantity $B_K$ means the kaon's binding energy. Details are given in the text.}
\label{Fig_ImageOfPict}
\end{figure}

We calculate a so-called kaon's binding energy $B_K$ as an auxiliary quantity, which is obtained by subtracting the $NN$ energy from the $\bar{K}NN$ energy: 
\begin{equation}\label{BK}
\begin{split}
B_K \; \equiv \; - \langle \Phi_{\bar{K}NN}^{\theta_P} | \, \hat{H}_{\bar{K}NN}^{\theta_P} \, - \, \hat{H}_{NN}^{\theta_P} \, |\Phi_{\bar{K}NN}^{\theta_P}\rangle,  
\end{split}
\end{equation}
where the term $\hat{H}_{NN}^{\theta_P}$ is a complex-scaled Hamiltonian for the $NN$ system. The $NN$ Hamiltonian is given as $\hat{H}_{NN}\,=\,\hat{\bm{p}}_1^2/2\mu_{NN}+\hat{V}_{NN}$.  
Using the kaon's binding energy, we estimate the $\bar{K}N$ energy with two ansatz based on two extreme concepts; 
\begin{equation}\label{E_KN}
\begin{split}
\sqrt{s}_{\bar{K}N} \; = \; \left\{
\begin{array}{lcl}
M_N + m_K - B_K   & \cdots & {\rm Field \; picture} \\
M_N + m_K - B_K/2 & \cdots & {\rm Particle \; picture}
\end{array}
\right. ,
\end{split}
\end{equation}
where $M_N$ and $m_K$ are nucleon and anti-kaon masses, respectively. 
On one ansatz, we consider the anti-kaon as a field which carries the kaon's binding energy. (See the left panel of Fig. \ref{Fig_ImageOfPict}.) Since the anti-kaon with the energy $\omega_K=m_K-B_K$ interacts with each nucleon here, the $\bar{K}N$ energy $\sqrt{s}_{\bar{K}N}$ is equal to $M_N+\omega_K$, namely $M_N + m_K - B_K$, with static approximation applied to nucleons.  On the other ansatz, we treat the anti-kaon as a particle. Since the anti-kaon is bound by two nucleons and the kaon's binding energy is provided by them, the binding energy per a $\bar{K}N$ bond should be  a half of $B_K$. (See the right panel of Fig. \ref{Fig_ImageOfPict}.) Therefore, the energy of a $\bar{K}N$ pair is equal to $M_N + m_K - B_K/2$. Hereafter, we denote the first ansatz as {\it Field picture} and the latter ansatz as {\it Particle picture}. For the convenience, we refer the $\bar{K}N$ energy measured from the $\bar{K}N$ threshold; $E(KN) \equiv \sqrt{s}_{\bar{K}N} - M_N - m_K$. 

When bound and resonant states of the $\bar{K}NN$ system are considered with such energy-dependent potentials, the self-consistency for the $\bar{K}N$ energy has to be taken into account as explained in the former study \cite{Kpp:DHW}. The $\bar{K}N$ energy set in the effective $\bar{K}N$ potential should finally coincide with that estimated with the obtained wave function by following the above-mentioned ansatz. It is noted that the self-consistency is realized for the {\it complex} $\bar{K}N$ energy in the current study. We, here, treat a resonant $\bar{K}NN$ state as a Gamow state with the correct boundary condition. Since the pole energy on the complex-energy plane is explicitly considered, the $\bar{K}N$ energy is treated as a complex value. On the other hand, such a self-consistency is considered only for the real energy in the former study with a variational approach \cite{Kpp:DHW}, since the $\bar{K}NN$ is treated within a bound-state approximation having a real binding energy.

\section{Nature of ccCSM+Feshbach method on $\bar{K}N$-$\pi Y$ system} \label{Sec:Result-2body}

Before the investigation of the $K^-pp$ system, we study the nature of our method on the two-body $\bar{K}N$-$\pi Y$ system. As the $\bar{K}N$(-$\pi Y$) potential $V_{\bar{K}N, \pi Y (I)}$ shown in Eq. (\ref{Ueff_KN}), which is the origin of the effective $\bar{K}N$ potential, we use a chiral SU(3)-based potential that was proposed in our previous study \cite{ccCSM-KN_NPA}. Our $\bar{K}N$(-$\pi Y$) potential is energy-dependent and is given with a single-range Gaussian form in the coordinate space. A non-relativistic version of this potential, called {\it NRv2c}, are employed in this section.

\subsection{Test calculation of ccCSM+Feshbach projection on a two-body $\bar{K}N$-$\pi Y$ system}

\begin{table}[t]
\caption{Quantities of an $I=0$ $\bar{K}N$-$\pi\Sigma$ system calculated with ccCSM and ccCSM+Feshbach. The quantities of 
$a_{\bar{K}N\,(I=0)}$ and $a_{\pi\Sigma\,(I=0)}$ are $\bar{K}N$ and $\pi\Sigma$ scattering lengths, respectively. The values ($-B(\bar{K}N)$, $-\Gamma/2$) indicate a resonance position of the system on the complex energy plane. $\sqrt{\langle r^2 \rangle}_{\bar{K}N}$ and $\sqrt{\langle r^2 \rangle}_{\pi\Sigma}$ are a meson-baryon mean distance of each component of the resonant state. Here, energies and lengths are given in units of MeV and fm, respectively.}
\label{Tab_ccCSM+F_test}
\centering
\begin{tabular}{l@{\hspace{2.0cm}}c@{\hspace{2.0cm}}c@{\hspace{1.5cm}}c}
\hline \hline
          & ccCSM  & \multicolumn{2}{c}{ccCSM+Feshbach} \\
$P$ space & $\bar{K}N, \; \pi\Sigma$  & $\bar{K}N$ & $\pi\Sigma$\\ 
\hline
$a_{\bar{K}N\,(I=0)}$  & $-1.700+0.681i$ & $-1.700+0.681i$ & ---\\
$a_{\pi\Sigma\,(I=0)}$ & $0.724$           & ---           & $0.724$\\
\hline
$B(\bar{K}N)$ & 17.150 & 17.156 & 17.415 \\
$\Gamma/2$    & 16.608 & 16.611 & 16.346 \\
%\hline
$\sqrt{\langle r^2 \rangle}_{\bar{K}N}$  & $1.280-0.403i$ & $1.280-0.403i$ & --- \\
$\sqrt{\langle r^2 \rangle}_{\pi\Sigma}$ & $0.233+0.930i$ &  --- & $0.234+0.932i$\\
%$D(\bar{K}N+\pi\Sigma)$ & $1.415-0.345i$ & --- & --- \\
\hline \hline
\end{tabular}
\end{table}

First, we examine the ccCSM+Feshbach projection method on a two-body $\bar{K}N$-$\pi Y$ system. Table \ref{Tab_ccCSM+F_test} shows the results of scattering and resonance properties of $I=0$ channel obtained with both methods of ccCSM and ccCSM+Feshbach. Here, the test calculation is performed with an energy-dependent potential (NRv2c). In the ccCSM both of $\bar{K}N$ and $\pi\Sigma$ components are explicitly treated as the model space ($P$ space), while in the ccCSM+Feshbach one component is set to be $P$ space and the other component is considered as $Q$ space to be eliminated. In the two-body case, the calculation of the ccCSM+Feshbach via an effective two-body potential is completely equivalent to that of the ccCSM treating all channels explicitly. In principle, results of both calculations should agree with each other. However, the $Q$-space Green function used in ccCSM+Feshbach is approximately represented with a finite number of Gaussian basis as explained in the section \ref{Method_Feshbach}. Under this approximation, the ccCSM+Feshbach is confirmed to reproduce quite well the ccCSM results of both scattering lengths and resonance properties. 

The scattering lengths are calculated with the CS-WF method which is a method to solve scattering problems with help of ccCSM as explained in the section 2.3 of our previous paper \cite{ccCSM-KN_NPA}. We plug the effective potential derived with ccCSM+Feshbach into the CS-WF method with a single $\bar{K}N$/$\pi\Sigma$ channel. It is noted that the scattering amplitudes of these components are confirmed to be identical between the two methods in wide energy region of $-200$ MeV to 50 MeV measured from the $\bar{K}N$ threshold. Furthermore, in the $I=1$ case that $\pi\Lambda$ channel is additionally coupled with $\bar{K}N$ and $\pi\Sigma$ channels, the ccCSM+Feshbach reproduces all the $\bar{K}N$, $\pi\Sigma$ and $\pi\Lambda$ scattering amplitudes obtained with the ccCSM. 

In the calculation of a resonance pole, the self-consistency for the $\bar{K}N$ energy is needed to be taken into account in both methods. As shown in Table \ref{Tab_ccCSM+F_test}, the pole position obtained self-consistently with the ccCSM+Feshbach is found to agree with that obtained with the ccCSM, whichever of $\bar{K}N$ and $\pi\Sigma$ channels is chosen as the $P$ space. The meson-baryon mean distance in $\bar{K}N$ and $\pi\Sigma$ components also coincides in both methods, when the normalization of each component in the ccCSM is appropriately considered; 
\begin{equation}\label{Wfunc-KNN}
\begin{split}
\langle r^2 \rangle_{MB} \; \equiv \; \langle \tilde{\phi}^\theta_{MB} | \, \hat{r}^2_\theta \, | \phi^\theta_{MB} \rangle \, / \, \langle \tilde{\phi}^\theta_{MB} | \phi^\theta_{MB} \rangle, 
\end{split}
\end{equation}
where $\phi^\theta_{MB}$ is a complex-scaled wave function of the $MB$ component of the resonant state and $\hat{r}^2_\theta$ indicates the complex-scaled operator of the meson-baryon distance. 
 
The resonance pole given in Table \ref{Tab_ccCSM+F_test} is the higher pole of the double pole obtained with our energy-dependent potential. It should be noted that the ccCSM+Feshbach reproduces well the result of the ccCSM also for the other pole, namely the lower pole \cite{ccCSM_DP:Dote}. The energies ($B(\bar{K}N)$, $\Gamma/2$) are (37.954, 135.943) MeV and (38.134, 136.169) MeV when the $\bar{K}N$ and $\pi\Sigma$ channels are selected to be the $P$ space in the ccCSM+Feshbach calculation, respectively. Those values agree well with ($B(\bar{K}N)$, $\Gamma/2$)=(38.128, 136.166) MeV obtained by the ccCSM.

\subsection{Dependence of two kinds of the scaling angles, $\theta_P$ and $\theta_Q$}

\begin{table}[t]
\caption{Dependence of the $\Lambda^*$ eigen energy on the scaling angles $\theta_P$ and $\theta_Q$. In the upper (lower) table, the scaling angle $\theta_Q$ ($\theta_P$) is varied with $\theta_P$ ($\theta_Q$) fixed to 30$^\circ$. ``NF'' means that no solutions are found below the $\bar{K}N$ threshold. NRv2 potential ($f_\pi=110$) is employed. Energies ($B(\bar{K}N)$ and $\Gamma/2$) are given in unit of MeV. The unit of scaling angles is degree. }
\label{Tab_ccCSM+F_test2}
\centering
\begin{tabular}{l@{\hspace{1.5cm}}c@{\hspace{1.0cm}}c@{\hspace{1.0cm}}c@{\hspace{1.0cm}}c@{\hspace{1.0cm}}c@{\hspace{1.0cm}}c@{\hspace{0.4cm}}}
\hline \hline
$\theta_P=30$ &  &  &  &  &  & \\
$\theta_Q$    & $<$5 & 10 & 15 & 20 & 25 & 30 \\ 
\hline
$B(\bar{K}N)$ & NF & 17.0982 & 17.1535 & 17.1562 & 17.1563 & 17.1563 \\
$\Gamma/2$    & NF & 13.8896 & 16.4649 & 16.6041 & 16.6111 & 16.6113 \\
\hline \hline
\end{tabular}

\vspace{0.5cm}

\begin{tabular}{l@{\hspace{1.5cm}}c@{\hspace{1.0cm}}c@{\hspace{0.9cm}}c@{\hspace{0.9cm}}c@{\hspace{0.9cm}}c@{\hspace{0.9cm}}c@{\hspace{0.0cm}}}
\hline \hline
$\theta_P$    &  0 & 5 & 10 & 15 & 20 & 30  \\
$\theta_Q=30$ &  &  &  &  & &  \\ 
\hline
$B(\bar{K}N)$  & 17.1558 & 17.1558 & 17.1558 & 17.1558 & 17.1558 & 17.1563\\
$\Gamma/2$     & 16.6128 & 16.6128 & 16.6128 & 16.6128 & 16.6128 & 16.6113\\
\hline \hline
\end{tabular}
\end{table}

We have made further investigation on properties of the ccCSM+Feshbach method. As explained in Section \ref{Method}, there are two kinds of scaling angles, $\theta_P$ and $\theta_Q$, in the method. The scaling angle $\theta_Q$ is introduced when we construct an effective potential by eliminating $Q$-space components with Feshbach method in Eqs. (\ref{Fesh_GQ}) and (\ref{Fesh_Ueff}). The other scaling angle $\theta_P$ is used to find resonance states by means of the complex scaling method for the $P$-space Hamiltonian which involves the effective potential. Table \ref{Tab_ccCSM+F_test2} shows the dependence of the energy of the $\Lambda^*$ resonant state on those angles, where $\Lambda^*$ means the $I=0$ resonance of the $\bar{K}N$-$\pi\Sigma$ system. In the upper table, we investigate the $\theta_Q$ dependence by fixing the angle $\theta_P$. In principle, the resonance energy should be independent of the angle $\theta_Q$, since the complex-scaled Green function for the $Q$ space $G_Q^{\theta_Q}(E)$ is inversely transformed to be a non-scaled Green function, $G_Q (E) = U^{-1}(\theta_Q) G_Q^{\theta_Q}(E) U(\theta_Q)$, in the construction of the effective potential. (See Eq. (\ref{Fesh_Ueff})) Certainly, the resonance energy is confirmed to be stable for $\theta_Q > 15^\circ$. However, around $\theta_Q=15^\circ$ the resonance energy becomes unstable, and then cannot be obtained for small angles $\theta_Q < 5^\circ$. We consider that this is due to insufficient description of the $Q$-space Green function. At such small scaling angles, since the extended closure relation is not well approximated with finite numbers of Gaussian basis functions, the $Q$-space Green function is not correctly represented \cite{CSM:Myo2}. Also in the former study of the complex scaling method where the level density was analyzed \cite{CSM-LD:Suzuki}, it is shown that the Green function is stably described with a basis function expansion when the scaling angle is chosen to be sufficiently large. On the other hand, we check the $\theta_P$ dependence in the lower table where $\theta_Q$ is fixed.  It is confirmed that the resonant energy is completely stable for the scaling angle $\theta_P$. Even at $\theta_P=0^\circ$, namely no scaling, the same resonance energy is obtained. Thus, once the Green function for the $Q$ space is well represented with a sufficiently large scaling angle $\theta_Q$, we can obtain resonant states within $P$ space correctly, using any scaling angle $\theta_P$. 

\subsection{Treatment of a complex effective potential}

\begin{table}[t]
\caption{Perturbative treatment of the imaginary part of the effective potential. ``(Full)'' indicates the full treatment of the complex potential, while ``(Perturb)'' indicates the perturbative treatment of the imaginary part of the complex potential. The two potentials, NRv1 and NRv2, are examined with a parameter $f_\pi$ varied from 90 to 120 MeV. Energies are given in unit of MeV.}
\label{Tab_ccCSM+F_test3}
\centering
\begin{tabular}{l@{\hspace{1.5cm}}c@{\hspace{0.5cm}}c@{\hspace{0.5cm}}c@{\hspace{0.5cm}}c@{\hspace{1.5cm}}c@{\hspace{0.5cm}}c@{\hspace{0.5cm}}c@{\hspace{0.5cm}}c@{\hspace{1.0cm}}}
\hline \hline
Potential     & NRv2 &     &     &     & NRv1 &     &     &  \\
$f_\pi$       & 90   & 100 & 110 & 120 & 90   & 100 & 110 & 120 \\ 
\hline
(Full)          &  &  &  &  &  &  &  & \\
$B(\bar{K}N)$ & 15.1 & 17.0 & 17.1 & 16.6 & 15.2 & 17.7 & 18.4 & 18.1 \\
$\Gamma/2$    & 23.1 & 19.8 & 16.6 & 14.0 & 26.0 & 23.1 & 19.5 & 16.7 \\
\hline
(Perturb)       &  &  &  &  &  &  &  &  \\
$B(\bar{K}N)$ & 19.1 & 18.1 & 17.1 & 16.3 & 21.2 & 20.1 & 18.9 & 18.1 \\
$\Gamma/2$    & 21.6 & 18.9 & 16.8 & 15.2 & 22.7 & 19.9 & 17.7 & 16.1 \\
\hline \hline
\end{tabular}
\end{table}

Here, we examine a treatment of an effective potential. When the channels energetically lower than the specified state are eliminated as $Q$ space in Feshbach method, the effective potential for $P$ space is in general a complex potential. In our case, we consider the state located between the $\bar{K}N$ and $\pi Y$ threshold energies. Therefore, the effective potential for the $\bar{K}N$ channel should be a complex potential, when the lower channels, $\pi Y$, are treated as $Q$ space. In this article, such a complex potential is treated directly as it is, and the self-consistency for the energy is considered also with a complex energy. Table \ref{Tab_ccCSM+F_test3} shows the binding energy and decay width of the $\Lambda^*$ resonant state obtained with such a full treatment of the complex potential (denoted as ``(Full)''), where two versions of our energy-dependent potential and several $f_\pi$ values are examined. On the other hand, so far, the imaginary part of the complex potential has often been treated perturbatively \cite{AMDK, Kpp:DHW, Kpp:BGL, KKp:Enyo}. 

We estimate the effect of the perturbative treatment of the imaginary potential within our method as follows. First, we construct the effective potential $U^{eff}_{\bar{K}N(I=0)} (E_{\bar{K}N})$ by the Feshbach method as explained in Section \ref{Method_Feshbach}. With only the real part of the effective potential, we construct a Hamiltonian for the $\bar{K}N$ system; $\hat{H}'_{\bar{K}N} = \hat{T}_{\bar{K}N} + {\rm Re} \, \hat{U}^{eff}_{\bar{K}N(I=0)} (E_{\bar{K}N})$, where $\hat{T}_{\bar{K}N}$ means the operator of $\bar{K}N$ relative kinetic energy. We diagonalize the Hamiltonian $\hat{H}'_{\bar{K}N}$ with Gaussian basis functions as usual, namely without the complex scaling. Since eigen energies of the Hamiltonian $\hat{H}'_{\bar{K}N}$ are real value, we consider the self-consistency for the $\bar{K}N$ energy within  real values. In other words, we set a real-valued energy in the effective potential. After we obtain a self-consistent solution with a binding energy $B(\bar{K}N)$, we estimate the half decay width $\Gamma/2$ by calculating the expectation value of the imaginary potential ${\rm Im} \, \hat{U}^{eff}_{\bar{K}N(I=0)}$ with the eigen wave function. 
%Note that the imaginary potential is attributed purely to the $\pi Y$-channels elimination when the $\bar{K}N$ energy set in the effective potential is a real value. 

The result of the perturbative treatment of the imaginary potential is shown in Table \ref{Tab_ccCSM+F_test3} (denoted as ``(Perturb)''). Compared with the result of full treatment as mentioned before, it is found that when the decay width is calculated to be small with the full treatment, the perturbative treatment gives a binding energy similar to that of the full treatment. However, in case that the large decay width is obtained in the full treatment, there is large difference among two treatments. In particular, in such a case the perturbative treatment tends to give a large binding energy compared with the full treatment. Therefore, the imaginary potential is found to give repulsive contribution to the binding energy when it is included explicitly in the calculation.

\section{Results of $K^-pp$ system with ccCSM+Feshbach method} \label{Sec:Result-3body}

In the section, results of the three-body system $K^-pp$ with the ccCSM+Feshbach method are shown. In the three-body calculation, the central part of Av18 potential \cite{Av18} is employed as a $NN$ potential $\hat{V}_{NN}$ which appears in Eq. (\ref{H_KNN}). As the $\bar{K}N$(-$\pi Y$) potential $V_{\bar{K}N, \pi Y (I)}$ in Eq. (\ref{Ueff_KN}), an energy-independent potential \cite{AY_2002} is used for a test calculation, and two kinds of non-relativistic version of our energy-dependent potential, called {\it NRv1c} and {\it NRv2c} \cite{ccCSM-KN_NPA}, are examined.  

We should comment on the applicability of the ccCSM+Feshbach method. As explained in the section \ref{Method_Feshbach}, the effective potential generated with the present method in Eq. (\ref{Ueff_KN}) generally has an energy dependence due to the channel elimination. In principle, such an energy dependence of the potential violates the three-body unitarity, when it is applied to the three-body $\bar{K}NN$-$\pi YN$ coupled-channel system. However, we expect that the $\bar{K}NN$ calculation with the ccCSM+Feshbach method is a good approximation with small unitarity violation, when the resonance pole is located close to the $\bar{K}NN$ threshold. 

We remark on the scaling angles $\theta_P$ and $\theta_Q$. As shown in the previous section, the result is found to be independent of the angle $\theta_P$ in the two-body case, if the angle $\theta_Q$ is sufficiently large to represent the $Q$-space Green function.  Therefore, in the three-body calculation mentioned hereafter, we take a common angle $\theta$ for the $\theta_P$ and $\theta_Q$; $\theta_P=\theta_Q\equiv\theta$.\footnote{As shown in Eq. (\ref{Ueff_KN}), the complex-scaling operator $U(\theta_Q)$ remains in the effective $\bar{K}N$ potential $U^{eff}_{\bar{K}N(I)}$. When the Hamiltonian for the $\bar{K}NN$, $\hat{H}_{\bar{K}NN}$ given in Eq. (\ref{H_KNN}), is complex-scaled with the operator $U(\theta_P)$, the effective potential is also transformed as $U(\theta_P) \, U^{eff}_{\bar{K}N(I)} \, U^{-1}(\theta_P)$. If both the scaling angles $\theta_P$ and $\theta_Q$ are chosen to be equal as mentioned above, the operator $U(\theta_Q)$ included in the effective potential is cancelled out by the operator $U(\theta_P)$. With such a choice of the scaling angles, we can calculate matrix elements of complex-scaled potential with Gaussian basis functions as usual. Otherwise, since the operator like $U(\theta_Q)U^{-1}(\theta_P)=U(\theta_Q-\theta_P)$ remains in the effective potential, we need to consider the transformation of the basis functions with that operator. In other words, setting $\theta_Q=\theta_P$,  we are free from such transformation of the basis functions.} Similarly to the two-body calculation, results of the three-body calculation are confirmed to be independent of the scaling angle $\theta$.

\subsection{Comparison of ``Field picture'' and ``Particle picture'' in an energy-independent potential case} \label{Sec:SC-AY}

\begin{figure}[t]
\centering\includegraphics[width=3.2in]{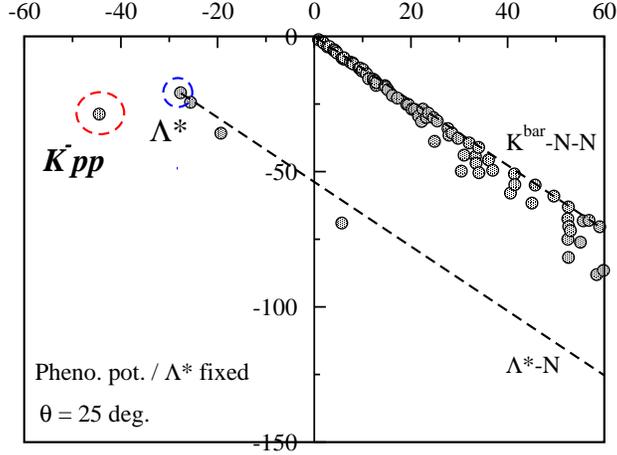}
\caption{Eigenvalue distribution of $\bar{K}NN$ system on the complex energy plane which is calculated with ccCSM+Feshbach method using a phenomenological potential \cite{AY_2002}. The $\bar{K}N$ energy is fixed at $\Lambda(1405)$. ``$\Lambda^*$'' means the $\Lambda(1405)$ resonance. Horizontal and vertical axes correspond to the real and imaginary parts of the complex $\bar{K}NN$ energy ``$E(\bar{K}NN)$'' which is measured from $\bar{K}+N+N$ threshold energy, respectively. The unit of energy is given in MeV.  The scaling angle $\theta$ is taken to be $25^\circ$.}
\label{Fig_Zdist-AY_Lfix}
\end{figure}

To consider the three-body system of $K^-pp$,  we investigate how the self-consistency for the $\bar{K}N$ energy is accomplished with Field and Particle pictures which are explained in the section \ref{Sec:SC}. For the simplicity, an energy-independent $\bar{K}N$(-$\pi Y$) potential, which is phenomenologically constructed \cite{AY_2002}, is here employed.

First, Fig. \ref{Fig_Zdist-AY_Lfix} shows the distribution of complex eigenvalues obtained with ccCSM+Feshbach method, when the $\bar{K}N$ energy is fixed at that of $\Lambda(1405)$. In this condition, the $\bar{K}N$ energy set in the $\bar{K}N$ effective potential is not consistent for the three-body $\bar{K}NN$ system, but it is consistent for the $I=0$ $\bar{K}N$ two-body system of $\Lambda(1405)$. In the figure, the origin corresponds to the $\bar{K}$-$N$-$N$ three-body threshold. Since eigenvalues of scattering continuum states are known to appear along the so-called $2\theta$ line in the complex scaling method, the eigenvalues along a line running from the origin indicate the $\bar{K}$-$N$-$N$ scattering continuum states. There is another line which starts from $E(\bar{K}NN)=(-27.7,\,-20.4)$ MeV (marked with blue-dashed circle in the figure). This energy is almost equal to the $\Lambda(1405)$ energy of $E(\bar{K}N)=(-28.2,\,-20.1)$ MeV which is set in the effective $\bar{K}N$ potential. Therefore, eigenvalues along this line indicate $\Lambda(1405)$-$N$ scattering continuum states. There is an eigenvalue isolated from two energy lines mentioned above (marked with red-dashed circle in the figure). This state corresponds to a $\bar{K}NN$ resonance. In the case that the $\bar{K}N$ energy is fixed to the $\Lambda(1405)$, the $\bar{K}NN$ resonance is obtained to be $E(\bar{K}NN)=(-44.5,\,-28.7)$ MeV. Thus, the $\bar{K}NN$ resonance can be identified with the ccCSM+Feshbach method. 

\begin{figure}[t]
\includegraphics[width=3.0in]{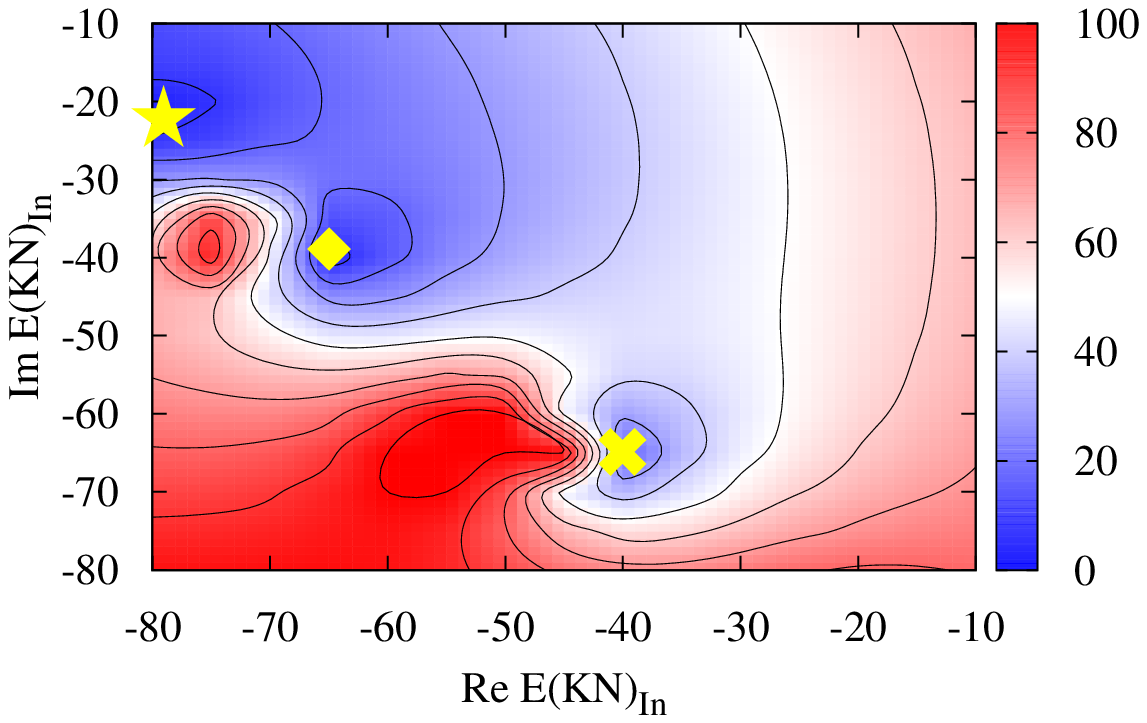}
\hspace{0.0cm}
\includegraphics[width=3.0in]{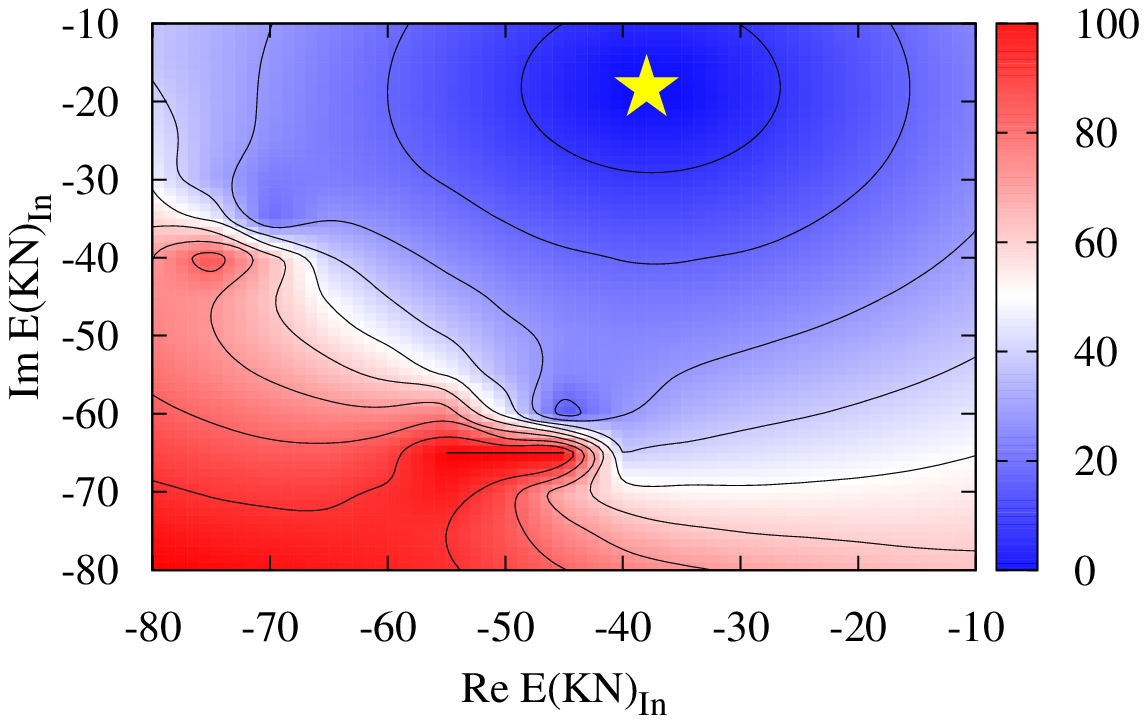}
\caption{Realization of the self-consistency for $\bar{K}N$ energy in case of a phenomenological potential at $\theta=25^\circ$. (Left) Field picture. (Right) Particle picture. The contour of the energy difference $|E(\bar{K}N)_{Cal}-E(\bar{K}N)_{In}|$ is depicted with respect to the complex value $E(\bar{K}N)_{In}$, where the variables $E(\bar{K}N)_{In}$ and $E(\bar{K}N)_{Cal}$ are the inputted and calculated $\bar{K}N$ energies, respectively. Detailed explanation is given in the text. The blue (red) color indicates smaller (larger) difference, which means the self-consistency is better (worse) achieved.  
%In the panels, the mesh size for the complex energy $E(\bar{K}N)_{In}$ is taken to be 5 MeV.
}
\label{Fig_SC-AY}
\end{figure}

Next, we calculate the pole energy of the $\bar{K}NN$ resonance, taking into account the self-consistency for the $\bar{K}N$ energy in the $\bar{K}NN$ three-body system. As explained in the section \ref{Sec:SC}, we consider that a self-consistent solution is obtained when the $\bar{K}N$ energy set in the effective potential ($E(KN)_{In}$) coincides with that calculated with the obtained wave function ($E(KN)_{Cal}$).  Fig. \ref{Fig_SC-AY} is a contour plot of the difference between the inputted and obtained $\bar{K}N$ energies, $|E(KN)_{Cal} - E(KN)_{In}|$, at each inputted complex $\bar{K}N$ energy. In case of Field picture, as shown in the left panel, three self-consistent solutions are found at $E(KN)_{In}=(-79,\,-22)$, $(-65,\,-39)$ and $(-43,\,-65)$ MeV which are marked with symbols of star, diamond and cross in the figure, respectively. However, when the stability of the solution for the scaling angle $\theta$ is examined, it is found that only the solution with $E(KN)_{In}=(-79,\,-22)$ MeV is stable, but that others are unstable. On the other hand, in case of Particle picture, as shown in the right panel of Fig. \ref{Fig_SC-AY} a single self-consistent solution is prominently found at $E(KN)_{In}=(-38,\,-18)$ MeV (marked with a symbol of star in the figure) and it is confirmed to be stable for the $\theta$ variation. 

\begin{table}[!h]
\caption{Self-consistent solutions with two ansatz, ``Field picture'' and ``Particle picture''. The $\bar{K}N$ energy of the obtained resonant state is given as $E(KN)$, and ``Re $E(KN)$'' and ``Im $E(KN)$'' are its real and imaginary parts, respectively. Values with parentheses mean the $\bar{K}N$ energy set in the effective $\bar{K}N$ potential. The binding energy and half decay width are given as ``$B(\bar{K}NN)$'' and ``$\Gamma/2$'', respectively. All energies are given in unit of MeV. ``$R(NN)$'' means a mean distance between two nucleons, and ``$R(\bar{K}{\text -}[NN])$'' means that between anti-kaon and center-of-mass of two nucleons. Unit of these lengths is fm. The last column lists a result of a different method ``ATMS'', which is quoted from Ref. \cite{Kpp:AY}.} 
\label{Tab_2an_AY}
\centering
\begin{tabular}{l@{\hspace{2cm}}c@{\hspace{1.5cm}}c@{\hspace{2.0cm}}c}
\hline \hline
Method          & \multicolumn{2}{l}{\hspace{1.0cm}ccCSM+Feshbach} & ATMS \cite{Kpp:AY}\\
$NN$ potential  & \multicolumn{2}{l}{\hspace{1.3cm}Av18 (Central)} & G3RS$^\dagger$\\
Ansatz          & Field pict. & Particle pict. &  ---  \\
\hline
Re $E(KN)$ & $-78.8$ \hspace{0.3cm} ($-79$) & $-37.7$ \hspace{0.3cm} ($-38$) & --- \\
Im $E(KN)$ & $-22.1$ \hspace{0.3cm} ($-22$) & $-18.0$ \hspace{0.3cm} ($-18$) & --- \\
\hline
$B(\bar{K}NN)$ & 48.9 & 45.8 & 48 \\
$\Gamma/2$     & 16.6 & 27.2 & 30.5 \\
%\hline
$R(NN)$           & $1.89-0.19i$ & $1.88-0.29i$ & 1.90 \\
$R(\bar{K}{\text -}[NN])$ & $1.27-0.14i$ & $1.25-0.21i$ & 1.35 \\
\hline \hline
\end{tabular}
\flushright
{\footnotesize $^\dagger$$^1E$ potential is commonly applied to the $^1O$ state. See the text.}
\end{table}

Details of the self-consistent solutions with ccCSM+Feshbach method using two pictures are given in Table \ref{Tab_2an_AY}. The binding energy of $\bar{K}NN$ ($B(\bar{K}NN)$), which is a real part of the resonance-pole energy, is not so different between two pictures. However, the half decay width ($\Gamma/2$), which is a imaginary part of the resonance-pole energy, is much different between them. Field picture gives half times smaller decay width than Particle picture. The spatial configuration is almost the same in those pictures, as indicated by the mean distance between two nucleons ($R(NN)$) and that between anti-kaon and center-of-mass of two nucleons ($R(\bar{K}{\text -}[NN])$). 

In the last column of the table, the result of an earlier study with ATMS method using the same $\bar{K}N$ potential \cite{Kpp:AY} is listed, for comparison with the present result. Among the two ansatz, Particle picture provides apparently the decay width close to that of the ATMS result. Note that they use a different $NN$ potential from Av18 central potential that we use in the current study. In addition, they apply the $^1E$ channel of the $NN$ potential commonly to the $^1O$ state. Since the $^1E$ potential is more attractive than the $^1O$ potential, their calculation is expected to give slightly deeper binding than our calculation. Taking into account such a difference on $NN$ potentials, Particle picture is considered to give the binding energy consistent to the ATMS result rather than Field picture. Thus, in a case of an energy-independent potential \cite{AY_2002}, our calculation of ccCSM+Feshbach method with {\it Particle picture} is found to give similar result of the former study with ATMS method. 

\subsection{The three-body $K^-pp$ system calculated with chiral SU(3)-based potentials}

\begin{table}[t]
\caption{Summary of self-consistent solutions of the $\bar{K}NN$ system with chiral SU(3)-based potentials, NRv1c and NRv2c. ``NF'' means that no $\theta$-stable solutions are found. All energies are given in unit of MeV.}
\label{Tab_sum_NRv12}
\centering
\begin{tabular}{l@{\hspace{1.5cm}}
c@{\hspace{0.5cm}}c@{\hspace{0.5cm}}c@{\hspace{0.5cm}}c@{\hspace{2.0cm}}
c@{\hspace{0.5cm}}c@{\hspace{0.5cm}}c@{\hspace{0.5cm}}c}
\hline \hline
$\bar{K}N$ pot. & \multicolumn{4}{l}{\hspace{1.0cm}NRv2c} & \multicolumn{4}{l}{\hspace{1.0cm}NRv1c}  \\
$f_\pi$         & 90 & 100 & 110 & 120 & 90 & 100 & 110 & 120 \\
\hline 
%               & & & & & & & & \\
Field pict.    & & & & & & & & \\
$B(\bar{K}NN)$ & NF & 32.2 & 25.6 & 21.2 & NF & 42.1 & 33.2 & 27.5 \\
$\Gamma/2$     & NF & 16.1 & 11.6 & 9.0  & NF & 16.1 & 13.3 & 10.7 \\
\hline
%               & & & & & & & & \\
Particle pict. & & & & & & & & \\
$B(\bar{K}NN)$ & 30.4 & 29.9 & 27.3 & 24.7 & 32.9 & 33.6 & 31.0 & 28.3 \\
$\Gamma/2$     & 31.7 & 24.2 & 18.9 & 15.2 & 36.8 & 28.5 & 22.1 & 17.7 \\
\hline \hline
\end{tabular}
\end{table}

In this section, we investigate the $K^-pp$ system with the present method using a chiral SU(3)-based $\bar{K}N$(-$\pi Y$) potential. Here, two versions of non-relativistic $\bar{K}N$ potentials, NRv1c and NRv2c, are employed. It should be noted that those potentials themselves have an energy dependence due to the chiral dynamics. In other words, the original coupled-channel potentials are already an energy-dependent potential, before they are converted to effective single-channel potentials which involve an energy dependence due to the channel elimination by the Feshbach projection. We comment on the difference of the energy dependence between those potentials. In the NRv2c potential, the energy dependence is completely attributed to the chiral dynamics. On the other hand, in the NRv1c potential another energy dependence is additionally involved which comes from so-called the flux factor that gives weak energy dependence. (See Eqs. (7) and (8) in Ref. \cite{ccCSM-KN_NPA}.)

\begin{figure}[t]
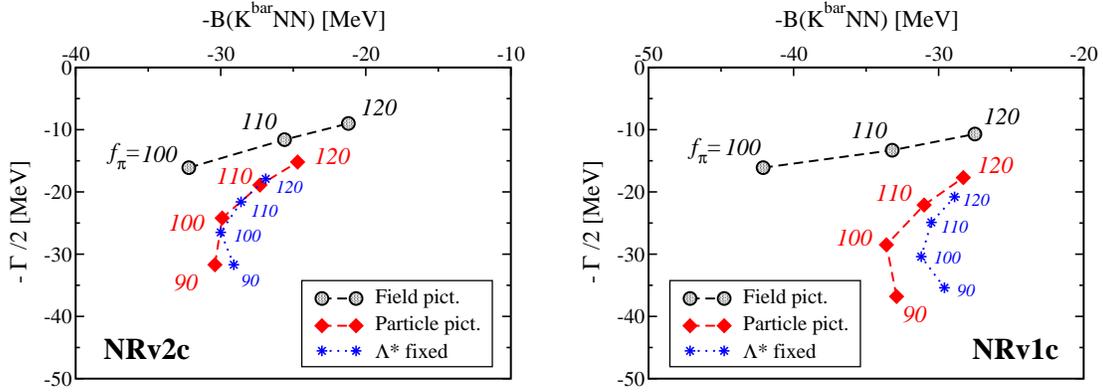

\centering
\includegraphics[width=2.7in]{Fig3-2_v3.eps}
\hspace{0.5cm}
\includegraphics[width=2.7in]{Fig3-1_v3.eps}
%%%call your figure name in the place "figurename.eps"
\caption{Pole energy of the self-consistent solution of the $\bar{K}NN$ resonance. Two versions of non-relativistic chiral SU(3)-based potential are employed: NRv2c potential (Left) and NRv1c potential (Right). A parameter $f_\pi$ of the potentials are varied from 90 to 120 MeV. Black-dashed line with filled circle and red-dashed line with filled diamond indicate Field and Particle pictures, respectively. The result obtained with ``$\Lambda$* fixed'' ansatz is also depicted with blue-dotted line with asterisk. 
%The scaling angle $\theta$ is set to be $30^\circ$.
}
\label{Fig_SC-NRv12}
\end{figure}

Table \ref{Tab_sum_NRv12} is the summary of the present calculation, which gives the binding energy and half decay width of the $\bar{KNN}$ resonance. In Fig. \ref{Fig_SC-NRv12} the pole position of the $\bar{KNN}$ resonance, ($-B(\bar{K}NN)$, $-\Gamma/2$), is depicted on the complex energy plane, when a parameter $f_\pi$ in the potentials is varied from 90 to 120 MeV. The left panel is the result obtained with NRv2c potential. From the figure, it is found that the $\bar{K}NN$ is bound more deeply and the decay width becomes wider, as the parameter $f_\pi$ decreases. However, the binding energy is not so large and it is 32 MeV at most. The binding energy is not so dependent on pictures for the $\bar{K}N$-energy self-consistency. On the other hand, the decay width depends strongly on the pictures. Similarly to the case of an energy-independent potential as mentioned in the previous section, the Field picture gives nearly half of the decay width, compared with the Particle picture. 
The result of NRv1c potential has a similar tendency to that of the NRv2c potential. In this potential, the binding energy is slightly larger compared with the NRv2c case, and it amounts to 42 MeV at $f_\pi=100$ MeV with Field picture. We consider that this is because the NRv1c potential is more attractive than the NRv2c potential, since the scattering amplitudes of the NRv1c potential show more attractive nature in the $\bar{K}N$ subthreshold region than those of the NRv2c potential as shown in Ref. \cite{ccCSM-KN_NPA}. It should be noted that in both potentials we cannot find any self-consistent solution which is stable for the $\theta$ variation at $f_\pi=90$ MeV with Field picture. As a reference, we show also the result that the $\bar{K}N$ energy set in the potential is fixed to the energy of $I=0$ $\bar{K}N$ resonance. (``$\Lambda^*$ fixed'' in the figure)  The result with this ansatz is quite similar to that with Particle picture. 
The binding energy $B(\bar{K}NN)$ and half decay width $\Gamma/2$ are summarized as 
\begin{equation}\label{PoleSum_NRv2c}
\begin{split}
(B(\bar{K}NN), \; \Gamma/2) \; = \; \left\{
%\begin{array}{lcl}
%(26.7 \pm 5.5, \quad 12.6 \pm 3.6) & \cdots & {\rm Field \; picture} \\
%(27.6 \pm 2.9, \quad 23.5 \pm 8.3) & \cdots & {\rm Particle \; picture}
%\end{array}
\begin{array}{lcl}
(21.2 \sim 32.2, \quad 9.0 \sim 16.1)  & \cdots & {\rm Field \; picture} \\
(24.7 \sim 30.4, \quad 15.2 \sim 31.7)  & \cdots & {\rm Particle \; picture}
\end{array}
\right. 
\end{split}
\end{equation}
in case of NRv2c, and 
\begin{equation}\label{PoleSum_NRv2c}
\begin{split}
(B(\bar{K}NN), \; \Gamma/2) \; = \; \left\{
%\begin{array}{lcl}
%(34.8 \pm 7.3, \quad 13.4 \pm 2.7) & \cdots & {\rm Field \; picture} \\
%(31.0 \pm 2.7, \quad 27.3 \pm 9.6) & \cdots & {\rm Particle \; picture}
%\end{array}
\begin{array}{lcl}
(27.5 \sim 42.1, \quad 10.7 \sim 16.1) & \cdots & {\rm Field \; picture} \\
(28.3 \sim 33.6, \quad 17.7 \sim 36.8) & \cdots & {\rm Particle \; picture}
\end{array}
\right. 
\end{split}
\end{equation}
in case of NRv1c. 

\begin{table}[t]
\caption{Self-consistent solutions of the $\bar{K}NN$ system with a chiral SU(3)-based potential, NRv2c ($f_\pi=110$ MeV). Distances ``$R(\bar{K}N,\,I=0)$'' and ``$R(\bar{K}N,\,I=1)$'' are a mean distance of a $\bar{K}N$ pair with isospin $I$ in the obtained $\bar{K}NN$ resonance. ``$\Lambda^*_{\rm vac.}$'' means a $I=0$ $\bar{K}N$-$\pi\Sigma$ resonance in vacuum. ``$E(\Lambda^*_{\rm vac.})$'' and  ``$R(\Lambda^*_{\rm vac.})$'' are a pole energy of the $\Lambda^*_{\rm vac.}$ and meson-baryon mean distance in the resonance, respectively. As for the meaning of other quantities, refer to the caption of Table \ref{Tab_2an_AY}. The last column ``Variational'' shows a result of a variational approach, which is quoted from Ref. \cite{Kpp:DHW}.}
\label{Tab_2an_NRv2}
\centering
\begin{tabular}{l@{\hspace{2.0cm}}c@{\hspace{1.0cm}}c@{\hspace{1.5cm}}c}
\hline \hline
Method          & \multicolumn{2}{l}{\hspace{1.2cm}ccCSM+Feshbach} & Variational \cite{Kpp:DHW}\\
$\bar{K}N$ pot. & \multicolumn{2}{l}{\hspace{2.2cm}NRv2c} & HW-HNJH  \\
Ansatz          & Field pict. & Particle pict. &  Particle pict.  \\
\hline
Re $E(KN)$ & $-54.9$ \hspace{0.3cm} ($-55$) & $-29.3$ \hspace{0.3cm} ($-29$) & $-22.6$ \hspace{0.3cm} ($-22.5$) \\
Im $E(KN)$ & $-17.3$ \hspace{0.3cm} ($-17$) & $-14.0$ \hspace{0.3cm} ($-14$) & --- \\
\hline
$B(\bar{K}NN)$ & 25.6 & 27.3 & 20.8 \\
$\Gamma/2$     & 11.6 & 18.9 & 29.2$^\dagger$ \\
%\hline
$R(NN)$                   & $2.21-0.24i$ & $2.11-0.29i$ & 2.15 \\
$R(\bar{K}{\text -}[NN])$ & $1.47-0.20i$ & $1.38-0.25i$ & --- \\
$R(\bar{K}N,\,I=0)$       & $1.61-0.21i$ & $1.51-0.27i$ & 1.73 \\
$R(\bar{K}N,\,I=1)$       & $2.34-0.28i$ & $2.22-0.32i$ & 2.26 \\
\hline
%$\Lambda^*_{E(KN)}$   & $-15.3 -8.6i$ & $-16.3 -14.4i$ & --- \\
%$R(\Lambda^*_{E(KN)})$  & $1.75 -0.44i$ & $1.49 -0.40i$ & --- \\
%\hline
$E(\Lambda^*_{\rm vac.})$   & \multicolumn{2}{l}{\hspace{1.5cm}$-17.2 -16.6i$ \cite{ccCSM-KN_NPA}}  & $-11.5 -21.9^\dagger i$\\
$R(\Lambda^*_{\rm vac.})$  & \multicolumn{2}{l}{\hspace{1.8cm}$1.42 -0.34i$ \cite{ccCSM-KN_NPA}}  & $1.86$\\
\hline \hline
\end{tabular}
\flushright
{\footnotesize $^\dagger$Perturbatively calculated with the wave function obtained in the variational approach.}
\end{table}

We have investigated the structure of the obtained $\bar{K}NN$ resonance. Table \ref{Tab_2an_NRv2} shows details of the self-consistent solution of the $\bar{K}NN$ calculated with the NRv2c potential with $f_\pi=110$ MeV, as a typical result. For the spatial configuration, several kinds of mean distances are given in the table. The mean distance between two nucleons ($R(NN)$) and that between anti-kaon and center-of-mass of two nucleons ($R(\bar{K}{\text -}[NN])$) are calculated with the obtained complex-scaled wave function of the $\bar{K}NN$ resonance $|\Phi_{\bar{K}NN}^\theta \rangle$ as 
\begin{equation}\label{Eq:Dist1}
\begin{split}
R(NN)^2 \equiv  \; 
\langle \Phi_{\bar{K}NN}^\theta | \, \hat{\bm x}_{1,\theta}^2 \, |\Phi_{\bar{K}NN}^\theta \rangle, \quad 
R(\bar{K}{\text -}[NN])^2 \equiv \; 
\langle \Phi_{\bar{K}NN}^\theta | \, \hat{\bm x}_{2,\theta}^2 \, |\Phi_{\bar{K}NN}^\theta \rangle, 
\end{split}
\end{equation}
where $\hat{\bm x}_{1,\theta}$ and $\hat{\bm x}_{2,\theta}$ indicate Jacobi-coordinate operators that are complex-scaled. Mean distance of a $\bar{K}N$ pair with isospin $I$ ($R(\bar{K}N,\,I)$) is calculated as 
\begin{equation}\label{Eq:Dist2}
\begin{split}
R(\bar{K}N,\,I)^2 \equiv  \; 
\langle \Phi_{\bar{K}NN}^\theta | \, \hat{\bm r}_{\bar{K}N,\theta}^2 \hat{P}_{\bar{K}N(I)}\, |\Phi_{\bar{K}NN}^\theta \rangle 
\; / \; 
\langle \Phi_{\bar{K}NN}^\theta | \, \hat{P}_{\bar{K}N(I)} \, |\Phi_{\bar{K}NN}^\theta \rangle, 
\end{split}
\end{equation}
where $\hat{P}_{\bar{K}N(I)}$ is an $\bar{K}N$ isospin projector. It is noted that two mean distances are obtained independently of $\theta$ \cite{CSM:Myo, CSM:Myo2, ccCSM-KN_NPA, ccCSM_DP:Dote}. In the complex scaling method, expectation values of distances are necessarily to be complex-valued since resonance states are treated as Gamow states. However, we refer to such complex-valued distances because we expect that they are useful guide for the spatial configuration of the resonant states, especially when the imaginary part of them is small compared with the real part.  

As a result of calculation of these mean distances, it is found that there is not so large difference between the results of two potentials. In both cases, the imaginary part of all mean distances is small compared to the real part. When we see the real part, the $NN$ mean distance is about 2.2 fm. For the comparison, the result of an earlier study with a variational calculation using a chiral SU(3)-based $\bar{K}N$ potential \cite{Kpp:DHW} is shown on the last column in the table. The $NN$ mean distance of the present study is found to be equal to that of the variational calculation. As mentioned in Ref. \cite{Kpp:DHW}, this $NN$ distance is almost identical to the mean distance between two nucleons in nuclear matter with normal density. As for the $\bar{K}N$ distance, the mean distance for the $I=0$ component is smaller than that for the $I=1$ component. This is due to the strong $\bar{K}N$ attraction in the $I=0$ channel. The variational calculation gives the similar result. However, the present values of the $\bar{K}N$ distance are smaller than those of the variational calculation. The same tendency has been found also in our previous study of the $\Lambda(1405)$ resonance treated as a two-body system of $I=0$ $\bar{K}N$-$\pi\Sigma$ \cite{ccCSM-KN_NPA}. We consider that such a difference is caused by the treatment of the $\bar{K}NN$ resonance: In the present study it is treated as a Gamow state, whereas it is treated as a bound state approximately in the variational study. By the way, compared with the case of in-vacuum where the $I=0$ $\bar{K}N$ pair forms a $\Lambda^*$ resonance, the mean distance of the $I=0$ $\bar{K}N$ pair is slightly larger in case of the $\bar{K}NN$ resonance. We consider that such a small elongation is due to the attraction from the other nucleon. 

\begin{table}[t]
\caption{Result of the $K^-pp$ calculation considering the SIDDHARTA data. NRv2c-SM1 and NRv2c-SM0 are the potentials where the SIDDHARTA result is taken into account, while NRv2c is the standard potential in the present study. $f_\pi=110$ MeV. The binding energy $B(\bar{K}NN)$ and half decay width $\Gamma/2$ are shown as ($B(\bar{K}NN)$, $\Gamma/2$) in the table. The energy is given in unit of MeV.}\label{Tab_K-pp_NRv2c-SM}
\centering
\begin{tabular}{l@{\hspace{1.0cm}}|@{\hspace{1.0cm}}
c@{\hspace{1.5cm}}c@{\hspace{1.5cm}}|@{\hspace{1.0cm}}c}
\hline \hline
               & NRv2c-SM1 & NRv2c-SM0   & NRv2c        \\
\hline
Field pict.    & (23.2, 15.5) & (26.2, 11.6) & (25.6, 11.6) \\
Particle pict. & (20.6, 23.6) & (28.1, 19.2) & (27.3, 18.9) \\
\hline \hline
\end{tabular}
\end{table}

As mentioned in the Introduction, SIDDHARTA collaboration reported a precise value of the $1s$-level energy shift of kaonic hydrogen atom \cite{Exp:SIDDHARTA}. Their result gives strong constraint to the $K^-p$ scattering length. Here, we consider the $K^-pp$ system, taking into account the SIDDHARTA data. We have constructed $\bar{K}N$(-$\pi Y$) potentials similarly to the NRv2c potential that is mainly used in the present study. Combining the $K^-p$ scattering length deduced from the SIDDHARTA data with the $I=0/I=1$ $\bar{K}N$ scattering length by the Martin's analysis, we have constructed two potentials: One is constrained with the SIDDHARTA $K^-p$ and Martin's $I=1$ $\bar{K}N$ scattering lengths (denoted as NRv2c-SM1), and the other is constrained with the SIDDHARTA $K^-p$ and Martin's $I=0$ $\bar{K}N$ scattering lengths (denoted as NRv2c-SM0). Detailed explanation of those potentials is given in Appendix \ref{App:NRv2c-SM}. We have carried out the ccCSM+Feshbach calculation of the $K^-pp$ system with those potentials. The obtained result is summarized in Table \ref{Tab_K-pp_NRv2c-SM}. In case of the NRv2c-SM1, the binding energy of $K^-pp$ becomes slightly smaller and the decay width increases, compared with the NRv2c. On the other hand, the NRv2c-SM0 gives almost the same result as the NRv2c. These results can be understood with the properties of those potentials. As explained in Appendix \ref{App:NRv2c-SM}, the NRv2c-SM1, in particular its $I=0$ part which gives dominant contribution in the $K^-pp$, is found to be slightly more attractive and absorptive than the NRv2c. The NRv2c-SM1 has quite the same nature as the NRv2c. Thus, both of NRv2c-SM1 and NRv2c-SM0 potentials result similar binding energy and decay width to those of the NRv2c potential. Therefore, we consider that the nature of $K^-pp$ system does not change so much in our treatment, even if the SIDDHARTA data is taken into account.

\subsection{$NN$/$\bar{K}N$ correlation density in the $K^-pp$ with the complex scaling method}

We investigate the spatial configuration of the $\bar{K}NN$ resonance in more detail. Similarly to a former study with a variational approach \cite{Kpp:DHW}, we calculate correlation density for $NN$ and $\bar{K}N$ pairs with the CSM wave function of the $\bar{K}NN$ resonance to visualize its structure. 

We give a brief explanation on the calculation of such densities, since they have to be calculated carefully in the complex scaling method. Here, we consider the case of the $NN$ correlation density, as an example. In the usual quantum mechanics, the expectation value of the $NN$-correlation-density operator, $\hat{\rho}_{NN}({\bm d}) \equiv \delta^3(\hat{\bm x}_1-{\bm d})$, is calculated. In the complex scaling method, a wave function is complex-scaled with the scaling operator $U(\theta)$. At the same time, an operator $\hat{O}$ is also complex-scaled as $\hat{O}_\theta = U(\theta) \hat{O} U^{-1}(\theta)$. With this transformation, the coordinate $\hat{\bm x}_n$ in the operator $\hat{O}$ is complex-scaled to be $\hat{\bm x}_n \exp(i\theta)$. Thus, the $NN$ correlation density $\rho_{NN}({\bm d})$ should be calculated with the CSM wave function as 
\begin{equation}\label{Eq:Dist2}
\begin{split}
\rho_{NN}({\bm d}) \quad &= \quad
\langle \tilde{\Phi}_{\bar{K}NN}^\theta | \, \hat{\rho}_{NN,\theta}({\bm d}) \, |\Phi_{\bar{K}NN}^\theta \rangle \\
&= \quad \int d{\bm x}_1 d{\bm x}_2 \; \Phi_{\bar{K}NN}^\theta ({\bm x}_1, {\bm x}_2) \, 
\delta^3({\bm x}_1 e^{i\theta}-{\bm d}) \, 
\Phi_{\bar{K}NN}^\theta ({\bm x}_1, {\bm x}_2)\\
&= \quad e^{-3i\theta} \int d{\bm x}_2 \; \Phi_{\bar{K}NN}^\theta ({\bm d}e^{-i\theta}, {\bm x}_2)^2 .
\end{split}
\end{equation}
In the same way, the isospin-separated $\bar{K}N$ correlation density $\rho_{\bar{K}N(I)}({\bm d})$ can be calculated, to begin with the operator $\hat{\rho}_{\bar{K}N(I)}({\bm d}) \equiv \delta^3(\hat{\bm r}_{\bar{K}N}-{\bm d}) \hat{P}_{\bar{K}N (I)}$ where $\hat{\bm r}_{\bar{K}N}$ is a $\bar{K}N$ relative coordinate operator and $\hat{P}_{\bar{K}N (I)}$ is a $\bar{K}N$ isospin projector. 

\begin{figure}[t]
\centering\includegraphics[width=3.0in]{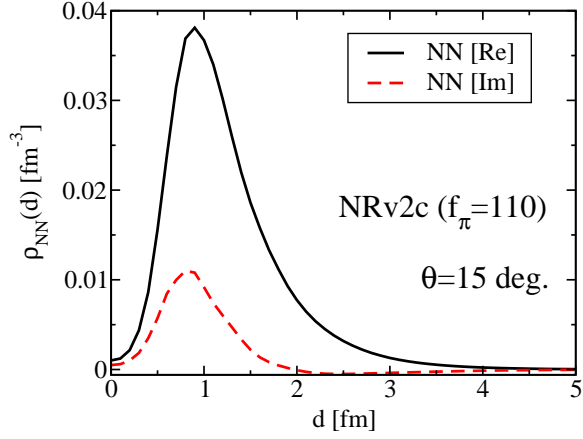}
\caption{$NN$ correlation density $\rho_{NN} (d)$ in the $\bar{K}NN$ resonance obtained with NRv2c potential ($f_\pi$=110 MeV) and Particle picture. The variable $d$ means a $NN$ relative distance. The real and imaginary parts of $\rho_{NN} (d)$ are drawn with black-solid and red-dashed lines, respectively. The density is normalized to unity. Length and density are given in unit of fm and fm$^{-3}$, respectively. The scaling angle is set to be 15$^\circ$.}
\label{Fig_CorrD-NN}
\end{figure}
 
\begin{figure}[t]
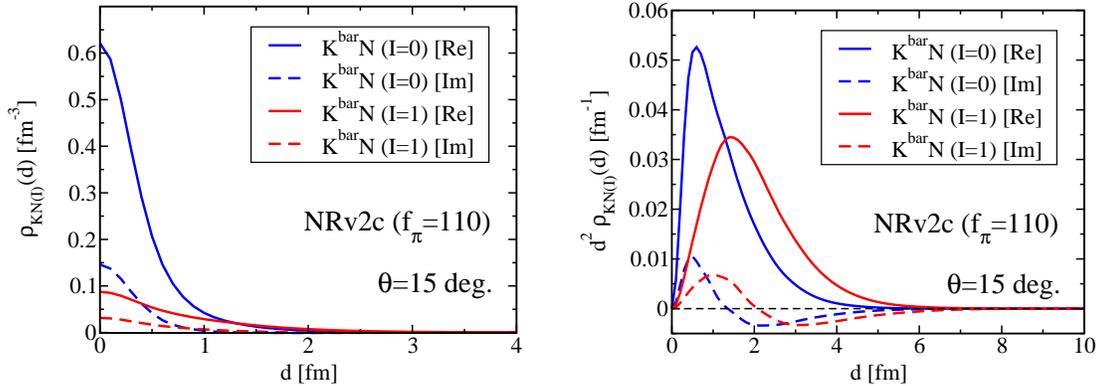

\centering
\includegraphics[width=2.7in]{Fig5-1_v2.eps}
\hspace{0.5cm}
\includegraphics[width=2.7in]{Fig5-2_v2.eps}
\caption{Isospin-separated $\bar{K}N$ correlation densities $\rho_{\bar{K}N(I)} (d)$ in the $\bar{K}NN$ resonance obtained with the same condition as Fig. \ref{Fig_CorrD-NN}. The variable $d$ means a $\bar{K}N$ relative distance. Densities of each isospin component are normalized to unity. The correlation densities of $I=0$ and $I=1$ components are depicted in blue and red colors, respectively. The real and imaginary parts of the density with each isospin component is drawn with solid and dashed lines, respectively. The right panel depicts the correlation densities multiplied by $d^2$ and the unit of vertical axis is fm$^{-1}$.}
\label{Fig_CorrD-KN}
\end{figure}

\begin{figure}[t]
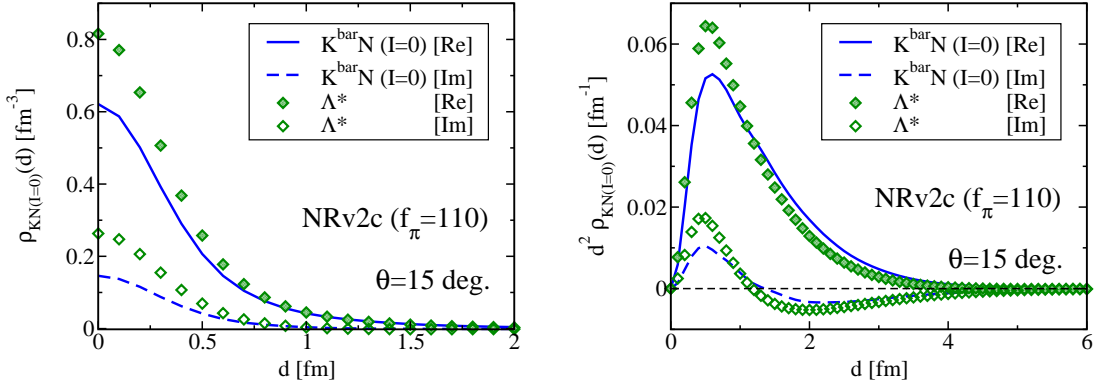

\includegraphics[width=2.7in]{Fig6-1_v2.eps}
\hspace{0.5cm}
\includegraphics[width=2.7in]{Fig6-2_v2.eps}
\caption{Comparison of $I=0$ $\bar{K}N$ correlation density in the $\bar{K}NN$ resonance and that of $\Lambda^*$ resonance. The $I=0$ $\bar{K}N$ correlation density of the $\Lambda^*$ resonance is shown with green diamonds, together with that in the $\bar{K}NN$ resonance shown in Fig. \ref{Fig_CorrD-KN}. The densities of both resonances are normalized to unity to be displayed. 
%The right panel depicts the correlation densities multiplied by $d^2$, where the variable $x$ means a $\bar{K}N$ relative distance.
}
\label{Fig_CorrD-KN_Lam}
\end{figure}

As explained above, we calculate the $NN$ and $\bar{K}N$ correlation densities with the obtained CSM wave function of the $\bar{K}NN$ resonance. Here, a typical result of the NRv2c potential ($f_\pi=110$ MeV) with Particle picture is displayed in Figs. \ref{Fig_CorrD-NN}-\ref{Fig_CorrD-KN_Lam}. The scaling angle is set to be a smaller angle of $15^\circ$ for the calculation of those densities. The pole position of the resonance is confirmed not to differ so much when the scaling angle is changed to $15^\circ$ from $30^\circ$ at which we have calculated so far: $(-B(\bar{K}NN), \, -\Gamma/2)$ are $(-27.4, \, -19.4)$ MeV at $\theta=15^\circ$, while they are $(-27.3, \, -18.9)$ MeV at $\theta=30^\circ$. 

Fig. \ref{Fig_CorrD-NN} shows the $NN$ correlation density. In a short distance, both the real and imaginary parts of the density is confirmed to be suppressed due to the strong repulsive core of the Av18 $NN$ potential. 
In Fig. \ref{Fig_CorrD-KN}, the $\bar{K}N$ correlation densities for the isospin 0 and 1 components are displayed. Both densities are normalized to unity for the comparison. In the figure, we can confirm directly the consequence of the strong $\bar{K}N$ attraction in the $I=0$ channel, as mentioned in the previous section. In other words, the $I=0$ $\bar{K}N$ component is found to distribute more compactly compared with the $I=1$ component. Furthermore, the $I=0$ $\bar{K}N$ correlation density is compared with the density of the $\Lambda^*$ resonance. The $\Lambda^*$ density is the $\bar{K}N$ density of two-body $\bar{K}N$ system with $I=0$ that is calculated with the ccCSM+Feshbach method using the same potential and the same scaling angle. As displayed in Fig. \ref{Fig_CorrD-KN_Lam}, both densities look rather similar to each other. Therefore, the present study with the ccCSM+Feshbach method also indicates that the $\Lambda^*$ resonance still survives in the $K^-pp$ resonance, as pointed out in the former study with a variational approach \cite{Kpp:DHW}.  

\section{Summary and future plan}

We have proposed a new method where the coupled-channel complex scaling method and the Feshbach projection are combined, and applied it (ccCSM+Feshbach method) to a three-body kaonic nucleus $K^-pp$. Originally $K^-pp$ is a $\bar{K}NN$-$\pi\Sigma N$-$\pi\Lambda N$ coupled-channel system, but has been effectively reduced to a single-channel problem of $\bar{K}NN$ by a channel elimination in the method. Recall that, since the ccCSM+Feshbach method is based on the complex scaling method, the $\bar{K}NN$ resonance has been regarded as a Gamow state with the correct boundary condition as a resonance. 

In ccCSM+Feshbach method, the extended closure relation (ECR), which is held in the complex scaling method (CSM), is essentially important. In fact, Green's function for $Q$-space (outer space of the model space $P$) needed to eliminate the $Q$ space, has been obtained easily from ECR, since the ECR is well realized with the $L^2$ Gaussian basis function only. First, we have tested the ccCSM+Feshbach method in a two-body $\bar{K}N$-$\pi Y$ coupled-channel system, and confirmed that it reproduces completely results of calculation treating all channels explicitly for both scattering and resonant problems. 

At the first application of the ccCSM+Feshbach method to the three-body system of $K^-pp$, we have examined an energy-independent $\bar{K}N$(-$\pi Y$) potential derived phenomenologically. Even when the original potential is energy-independent, the effective $\bar{K}N$ potential appeared in the ccCSM+Feshbach method
has an energy dependence as a result of channel elimination by the Feshbach projection, and the self-consistency for the complex $\bar{K}N$ energy has to be considered when we search for resonance states. We have tested two ansatz for the self-consistency, Field picture and Particle picture, and obtained a self-consistent solution successfully for each ansatz. We have found that the binding energy of $\bar{K}NN$ does not depend on the ansatz, while the decay width depends on the ansatz strongly: Field picture gives decay width so small as half of Particle picture. Compared with the result of an earlier study with ATMS method using the same potential \cite{Kpp:AY}, the result of Particle picture is found to be close to the past result; $(B(\bar{K}NN), \, \Gamma/2) = (45.8, \, 27.2)$ MeV. 

For more theoretical investigation of the $K^-pp$ system, we have used an energy-dependent $\bar{K}N$(-$\pi Y$) potential which was proposed in our previous work based on the chiral SU(3) theory. Two versions of non-relativistic potentials, NRv1c and NRv2c \cite{ccCSM-KN_NPA}, have been examined. Also for those energy-dependent potentials, self-consistent solutions of the $\bar{K}NN$ resonance with the ccCSM+Feshbach method have been found. Similarly to the case of the energy-independent potential, the decay width is rather small in Field picture compared with that in Particle picture. In case of NRv2c potential, the binding energy and half decay width of the $\bar{K}NN$ resonance are obtained to be $(21.2 \sim 32.2, \; 9.0 \sim 16.1)$ MeV with Field picture and $(24.7 \sim 30.4, \; 15.2 \sim 31.7)$ MeV with Particle picture. The NRv1c potential gives slightly larger binding energy. In addition, it is confirmed that the $\bar{K}NN$ resonance energy is not so varied, even if the precise value of $K^-p$ scattering length deduced from the SIDDHARTA experiment \cite{Exp:SIDDHARTA} is taken into account. 

As for the spatial configuration of $\bar{K}NN$ resonance, we have calculated $NN$ and $\bar{K}N$ correlation densities from the wave function, carefully following the procedure of the complex scaling method. Those densities are useful tool for intuitive understanding of the structure of $\bar{K}NN$ although they are given as complex values in the complex scaling method. The $NN$ correlation density is strongly suppressed at short distances as a result of the $NN$ repulsive core. The $NN$ mean distance is found to be about $2.2$ fm with a small imaginary part. This distance is almost equal to the $NN$ distance of a nuclear matter with the normal density. Concerning the $\bar{K}N$ correlation density, we have reconfirmed the survival of the $\Lambda^*$ resonance ($I=0$ $\bar{K}N$ resonance) in the $\bar{K}NN$ resonance, which was pointed out in a previous study with a variational approach \cite{Kpp:DHW}.  

Thus, through the present study with the ccCSM+Feshbach method, we have confirmed that the $K^-pp$ system is shallowly bound with a chiral SU(3)-based energy-dependent potential, as reported in earlier studies employing the same type of $\bar{K}N$ potential \cite{Kpp:DHW, Kpp:BGL, Kpp:IKS}. However, in case of NRv2c potential with Particle picture, we have always obtained another quasi self-consistent solution around the $\bar{K}N$ energy $E(\bar{K}N) \sim (-60, -60)$ MeV. (``quasi self-consistent'' means a local minimum for the quantity $|E(\bar{K}N)_{In} - E(\bar{K}N)_{Cal}|$ which is an indicator for the self-consistency as explained in the section \ref{Sec:SC-AY}.) In such quasi self-consistent solutions, a $\bar{K}NN$ resonance appears near the $\pi\Sigma N$ threshold with large decay width. Since $\Lambda(1405)$ has the double pole structure with the NRv2c potential as many studies with chiral SU(3) models \cite{ccCSM_DP:Dote}, we think that those quasi-consistent solutions are probably related to the lower pole of $\Lambda(1405)$. While, the solutions reported in the previous section must be related to the higher pole. In other words, $K^-pp$ is supposed to have the double pole structure as same as the $\Lambda(1405)$, as suggested in an earlier work with Faddeev-AGS approach \cite{Kpp-DP:IKS}. In order to have definite conclusion on this issue, we need more delicate calculation for the deeper pole, since it has large imaginary part. 

In this article, we have successfully obtained the solution of the $K^-pp$ resonance with the ccCSM+Feshbach method. However, since we have eliminated $\pi YN$ channel by the Feshbach projection, some of $\pi YN$ dynamics might be lost in the present study. Toward more decisive conclusion on the $K^-pp$ problem, we will carry out a coupled-channel three-body calculation with explicit $\pi YN$ channels. The coupled-channel calculation will clarify the detailed property of the $K^-pp$, such as the composition of the $K^-pp$ resonant state. On the experimental side, results of the $K^-pp$ search are going to be reported from two experimental groups at J-PARC (E15 \cite{Kpp-ex:JPARC-E15} and E27 \cite{Kpp-ex:JPARC-E27}). We hope that these experimental results will provide us with useful information of the $K^-pp$. 

Since extending application of the ccCSM+Feshbach method to four-body systems is straightforward, we can investigate rather easily four-body systems such as a kaonic nucleus $K^-ppn$ and a double kaonic nucleus $K^-K^-pp$, which have been investigated with a variational method \cite{Kpp:BGL} and Faddeev-Yakubovsky approach \cite{KKpp:Meada}. Generally, the method can be applied to various kinds of mesic nuclei which involve some decay modes, for example, mesic nuclei with $\eta$ \cite{eta:Nagahiro, eta:Mares}, $\eta'$ \cite{etap:Nagahiro}, $\omega$ mesons and $D$ meson in charm sector \cite{DNN}. Those are interesting systems and in the scope of our study with the ccCSM+Feshbach method.

\section*{Acknowledgment}

This work was developed through several activities (meetings  and workshops) at the J-PARC Branch of the KEK Theory Center. One of authors (A. D.) is thankful to Prof. T. Harada for fruitful discussion and to Prof. A. Ohnishi for his encouragement, and appreciates Dr. T. Hyodo's useful advice to improve this article. This work is supported by JSPS KAKENHI Grant Number 25400286 and partially by Grant Number 24105008. The calculation for this work was performed with High Performance Computing system (saho) at Research Center for Nuclear Physics (RCNP) in Osaka University.

\appendix

\section{Chiral SU(3)-based $\bar{K}N$(-$\pi Y$) potentials constrained with SIDDHARTA data}
\label{App:NRv2c-SM}

In this appendix, we explain details of our chiral SU(3)-based potential in which the SIDDHARTA data on kaonic hydrogen atom \cite{Exp:SIDDHARTA} is taken into account. 

We can deduce the $K^-p$ scattering length from the $1s$-level energy shift of kaonic hydrogen atom measured in the SIDDHARTA experiment. With help of the improved Deser-Trueman formula \cite{MDT:Meissner}, the $K^-p$ scattering length is obtained to be $a_{K^-p} = -0.65 + 0.81i$ fm \cite{ChU:IHW}. To determine the parameters of our potential in each isospin channel, we need the $I=0$ and $I=1$ $\bar{K}N$ scattering lengths individually, although the $K^-p$ scattering length is given by the average of the two isospin cases of $I=0$ and $I=1$; $a_{K^-p} = \{a_{\bar{K}N(I=0)}+a_{\bar{K}N(I=1)}\}/2$. Here, combining with the $\bar{K}N$ scattering length by the Martin's analysis of old data \cite{Exp:ADMartin}, we determine the $\bar{K}N$ scattering length of each isospin channel from the SIDDHARTA $K^-p$ scattering length in the following two cases: 
%\begin{eqnarray}
%&\hspace*{-1.5cm}\bullet& \hspace*{-0.8cm}a_{K^-p}{\rm :SIDDHARTA} \;\; -0.65 + 0.81i, \quad a_{\bar{K}N(I=1)}{\rm :Martin} \;\; 0.37 + 0.60i \; {\rm fm} \nonumber \\
%&& \hspace*{7cm} \Longrightarrow a_{\bar{K}N(I=0)}=-1.67+1.02i \; {\rm fm} \label{aKN_SM1}\\
%&\hspace*{-1.5cm}\bullet& \hspace*{-0.8cm}a_{K^-p}{\rm :SIDDHARTA}  \;\; -0.65 + 0.81i, \quad a_{\bar{K}N(I=0)}{\rm :Martin} \; -1.70 + 0.68i \; {\rm fm} \nonumber \\
%&& \hspace*{7cm} \Longrightarrow a_{\bar{K}N(I=1)}= 0.40+0.94i \; {\rm fm} \label{aKN_SM0}
%\end{eqnarray}

%\begin{table}[h]
%\centering
\vspace{0.5cm}
\begin{tabular}{l@{\hspace{0.5cm}}@{\hspace{1.5cm}}c@{\hspace{1.5cm}}c@{\hspace{1.5cm}}c}
\hline
  & $a_{K^-p}$ [fm]   & $a_{\bar{K}N(I=0)}$ [fm] & $a_{\bar{K}N(I=1)}$ [fm] \\
\hline
$\bullet$ Case 1  & SIDDHARTA       & $\Longrightarrow$ determined & Martin \\
  & $-0.65 + 0.81i$ & $-1.67+1.02i$     & $0.37 + 0.60i$ \\
\\
$\bullet$ Case 2  & SIDDHARTA       & Martin            & $\Longrightarrow$ determined\\
  & $-0.65 + 0.81i$ & $-1.70 + 0.68i$   & $0.40+0.94i$ \\
\hline
\end{tabular}
%\end{table}
\vspace{0.5cm}

\noindent
Note that the Martin's scattering length is given for each isospin channel and has been used to determine our potentials used in this article \cite{ccCSM-KN_NPA}. 

Here, we begin with our standard potential, NRv2c with $f_\pi=110$ MeV. To reproduce the above-mentioned $\bar{K}N$ scattering lengths, we adjust the range parameters of the NRv2c potential which are defined in Eq. (8) of Ref. \cite{ccCSM-KN_NPA}. We denote the potentials which are constrained with the $\bar{K}N$ scattering lengths of Case 1 and Case 2 in the above table as {\it NRv2c-SM1} and {\it NRv2c-SM0}, respectively. The range parameters of those potentials and $\bar{K}N$ scattering lengths calculated with them are summarized in Table \ref{Tab_NRv2c-SM}. We make two remarks: 1) Those potentials reproduce only the imaginary part of the scattering length in the $I=1$ channel, following the guideline of construction of NRv2c potential \cite{ccCSM-KN_NPA}. 2) As for the $I=0$ channel, the NRv2c-SM0 potential is the same as the NRv2c potential, since the range parameters of  the $I=0$ channel in both potentials are fixed to reproduce the Martin's $I=0$ value. (See Case 2 in the above table)

\begin{table}[b]
\caption{Properties of NRv2c-SM1 and NRv2c-SM0 potentials ($f_\pi=110$ MeV). Upper and lower tables are for the isospin $I=0$ and $I=1$ channels, respectively.  The range parameters of Gaussian form factor of the potential, $d^{(I=0,\,1)}_{\bar{K}N, \bar{K}N}$, $d^{(I=0,\,1)}_{\pi\Sigma, \pi\Sigma}$ and $d^{(I=1)}_{\bar{K}N, \pi\Lambda}$, are given in unit of fm. They are specified in the same as Tables 1 and 3 in our previous work \cite{ccCSM-KN_NPA}. The scattering length $a_{\bar{K}N(I)}$ is calculated with those potentials in each isospin channel. All lengths are in unit of fm. Complex values of $Z_H$ and $Z_L$ in the upper table are the complex energy of higher and lower poles of the $I=0$ $\bar{K}N$-$\pi\Sigma$ system, respectively. These energies are measured from $\bar{K}N$ threshold and given in unit of MeV.}
\label{Tab_NRv2c-SM}
\centering
\begin{tabular}{l@{\hspace{0.25cm}}|
c@{\hspace{0.5cm}}c@{\hspace{0.3cm}}|@{\hspace{0.3cm}}c@{\hspace{0.5cm}}|@{\hspace{0.3cm}}c@{\hspace{0.3cm}}
c}
\hline \hline
 $I=0$ & $d^{(I=0)}_{\bar{K}N, \bar{K}N}$ & $d^{(I=0)}_{\pi\Sigma, \pi\Sigma}$ & $a_{\bar{K}N(I=0)}$ & $Z_H$ & $Z_L$ \\
\hline
NRv2c-SM1 & 0.458 & 0.587 & $-1.670+1.028i$ & $-8.5-20.7i$ & $-31.9-108.5i$ \\
NRv2c(-SM0) & 0.438 & 0.636 & $-1.700+0.681i$ & $-17.2-16.6i$ & $-39.8-137.9i$ \\
\hline \hline
\end{tabular}

\vspace{0.3cm}

\begin{tabular}{l@{\hspace{0.5cm}}|@{\hspace{1.5cm}}
c@{\hspace{1.0cm}}c@{\hspace{1.0cm}}c@{\hspace{1.0cm}}|@{\hspace{1.0cm}}c@{\hspace{0.9cm}}}
\hline \hline
 $I=1$ & $d^{(I=1)}_{\bar{K}N, \bar{K}N}$ & $d^{(I=1)}_{\pi\Sigma, \pi\Sigma}$ & $d^{(I=1)}_{\bar{K}N, \pi\Lambda}$ & $a_{\bar{K}N(I=1)}$  \\
\hline
NRv2c-SM1 & 0.458 & 0.587 & 0.444 & $0.635+0.601i$ \\
NRv2c-SM0 & 0.438 & 0.636 & 0.363 & $0.690+0.937i$ \\
NRv2c     & 0.438 & 0.636 & 0.445 & $0.657+0.599i$ \\
\hline \hline
\end{tabular}
\end{table}

\begin{figure}[t]
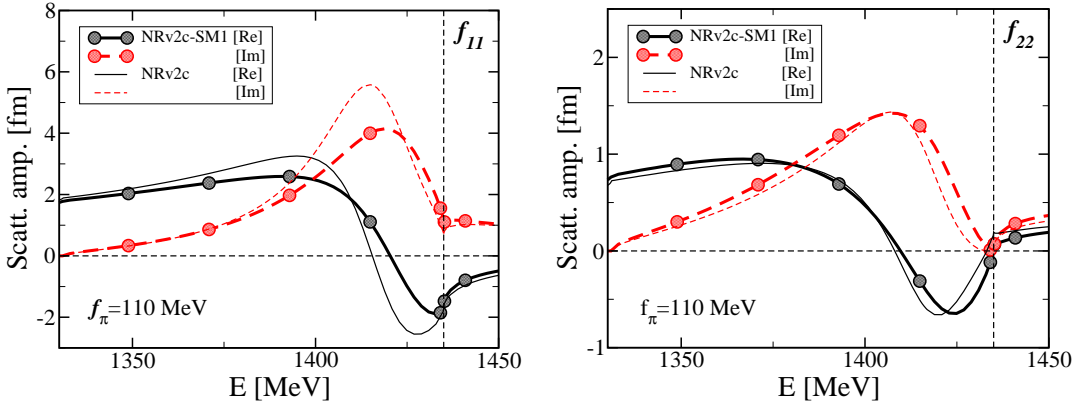

\centering
\includegraphics[width=2.7in]{FigA1-1_v2.eps}
\hspace{0.2cm}
\includegraphics[width=2.7in]{FigA1-2_v2.eps}
\caption{Comparison of $I=0$ scattering amplitudes calculated with NRv2c-SM1 and NRv2c potentials. (Left) $\bar{K}N$ scattering amplitude. (Right) $\pi\Sigma$ scattering amplitude. Bold line with filled circle (thin line) corresponds to the scattering amplitudes of NRv2c-SM1 (NRv2c) potential. Black-solid (red-dashed) line represents the real (imaginary) part of the scattering amplitude. The vertical dashed line indicates the $\bar{K}N$ threshold. 
}
\label{Fig_NRv2c-SM-I=0}
\end{figure}

We have investigated the properties of those potentials. The pole positions of 
the $I=0$ $\bar{K}N$-$\pi\Sigma$ resonances are given in the upper part of Table \ref{Tab_NRv2c-SM}. Those potentials have two poles, higher pole $Z_H$ and lower pole $Z_L$, in the $I=0$ channel, similarly to the NRv2c potential \cite{ccCSM_DP:Dote}. The resonant poles of the NRv2c-SM1 potential are less bound than those of the NRv2c potential. In the $I=0$ $\bar{K}N$ scattering amplitude, the resonance structure is certainly shifted to weaker binding side (namely, closer to the $\bar{K}N$ threshold), compared with the NRv2c potential. (See the left panel of Fig. \ref{Fig_NRv2c-SM-I=0}) In addition, the $I=0$ $\bar{K}N$ scattering length has large imaginary part in case of the NRv2c-SM1 potential. Reflecting this fact, the higher pole of the $I=0$ resonances, which is strongly coupled to the $\bar{K}N$ channel, has also large imaginary energy. By the way, the NRv2c-SM0 potential shows the same poles and scattering amplitudes as the NRv2c potential, since the $I=0$ part of both potentials are constrained with the Martin's value as explained above. Thus, it is found that our chiral SU(3)-based potential becomes less attractive and more absorptive in the $I=0$ channel, when the SIDDHARTA result is taken into consideration. 

\begin{figure}[t]
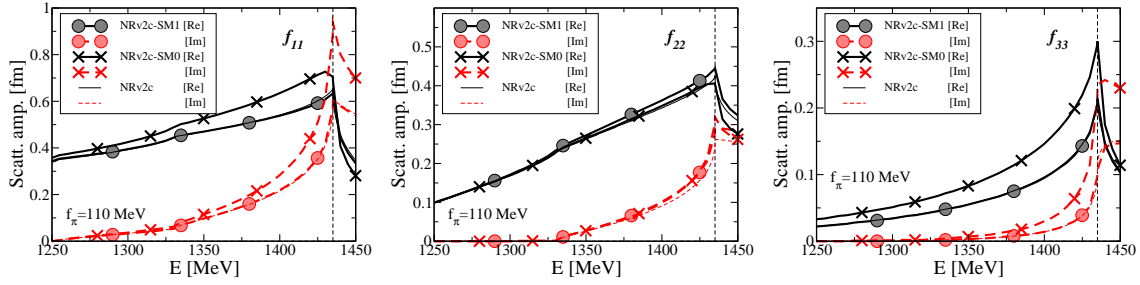

\includegraphics[width=1.9in]{FigA2-1_v2.eps}
\hspace{0.0cm}
\includegraphics[width=1.9in]{FigA2-2_v2.eps}
\hspace{0.0cm}
\includegraphics[width=1.9in]{FigA2-3_v2.eps}
\caption{Comparison of $I=1$ scattering amplitudes calculated with NRv2c-SM1, NRv2c-SM0 and NRv2c potentials. (Left) $\bar{K}N$ scattering amplitude. (Middle) $\pi\Sigma$ scattering amplitude. (Right) $\pi\Lambda$ scattering amplitude. Bold line with filled circles, bold line with crosses and thin line correspond to the scattering amplitudes of NRv2c-SM1, NRv2c-SM0 and NRv2c potentials, respectively. %Black-solid (red-dashed) line represents the real (imaginary) part of the scattering amplitude.
}
\label{Fig_NRv2c-SM-I=1}
\end{figure}

As for the $I=1$ channel, the NRv2c-SM0 potential has large imaginary part of the $\bar{K}N$ scattering length, compared with the NRv2c-SM1 and NRv2c potentials, although all three potentials give the similar real part of the $\bar{K}N$ scattering length. (Lower part of Table \ref{Tab_NRv2c-SM}) Fig. \ref{Fig_NRv2c-SM-I=1} depicts the $\bar{K}N$, $\pi\Sigma$ and $\pi\Lambda$ scattering amplitudes with $I=1$. The NRv2c-SM1 potential gives quite the same amplitudes as the NRv2c potential. (The amplitudes of both potentials are overlapped in the figure.) The $\bar{K}N$ amplitude of NRv2c-SM0 potential is slightly attractive than that of other potentials. However, the difference of scattering amplitudes between three potentials is tiny. Therefore, the property of our potential in the $I=1$ channel is found not to change so much with the SIDDHARTA result.

% can use a bibliography generated by BibTeX as a .bbl file
% BibTeX documentation can be easily obtained at:
% http://www.ctan.org/tex-archive/biblio/bibtex/contrib/doc/

%\bibliographystyle{ptephy}
%\bibliography{sample}
%
% once the .bbl file has been generated then place the text in your article.

\vfill\pagebreak

%\section*{References}

\end{document}